\pdfoutput=1 
\documentclass[cernpreprint,texlive=2011,txfonts,UKenglish]{latex/atlasdoc} 
\usepackage{latex/atlaspackage} 
\usepackage{latex/atlasphysics} 
\usepackage{latex/atlasbiblatex}
\usepackage{bigdelim}
\usepackage{multirow}

\newcommand*{\ttZ}{\ensuremath{\ttbar Z}\xspace}
\newcommand*{\ttW}{\ensuremath{\ttbar W}\xspace}
\newcommand*{\ttV}{\ensuremath{\ttbar V}\xspace}
\newcommand*{\ttH}{\ensuremath{\ttbar H}\xspace}
\newcommand*{\OS}{opposite-sign}
\newcommand*{\SSL}{same-sign dilepton}
\newcommand*{\OSL}{\OS\ dilepton}
\newcommand*{\TL}{trilepton}
\newcommand*{\FL}{tetralepton}
\newcommand*{\SSLC}{\SSL\ channel}
\newcommand*{\OSLC}{\OSL\ channel}
\newcommand*{\TLC}{\TL\ channel}
\newcommand*{\FLC}{\FL\ channel}
\newcommand*{\OSLSRA}{\mbox{2$\ell$-noZ-4j}}
\newcommand*{\OSLSRB}{\mbox{2$\ell$-noZ-5j}}
\newcommand*{\OSLSRC}{\mbox{2$\ell$-Z-5j}}
\newcommand*{\SSLSRA}{\mbox{$2e$-SS}}
\newcommand*{\SSLSRB}{\mbox{$e\mu$-SS}}
\newcommand*{\SSLSRC}{\mbox{$2\mu$-SS}}
\newcommand*{\TLSRA}{\mbox{3$\ell$-Z-1b4j}}
\newcommand*{\TLSRB}{\mbox{3$\ell$-Z-2b3j}}
\newcommand*{\TLSRC}{\mbox{3$\ell$-Z-2b4j}}
\newcommand*{\TLSRD}{\mbox{3$\ell$-noZ-2b}}
\newcommand*{\FLSRA}{\mbox{4$\ell$-DF-0b}}
\newcommand*{\FLSRB}{\mbox{4$\ell$-DF-1b}}
\newcommand*{\FLSRC}{\mbox{4$\ell$-DF-2b}}
\newcommand*{\FLSRD}{\mbox{4$\ell$-SF-1b}}
\newcommand*{\FLSRE}{\mbox{4$\ell$-SF-2b}}
\newcommand*{\OSLCRA}{\mbox{2$\ell$-noZ-3j}}
\newcommand*{\OSLCRB}{\mbox{2$\ell$-Z-3j}}
\newcommand*{\OSLCRC}{\mbox{2$\ell$-Z-4j}}
\newcommand*{\TLCR}{\mbox{3$\ell$-Z-0b3j}}
\newcommand*{\FLCR}{\mbox{4$\ell$-ZZ}}
\newcommand*{\rone}{\mbox{2$\ell$-noZ}}
\newcommand*{\rtwo}{\mbox{2$\ell$-Z}}
\newcommand*{\threezveto}{\mbox{3$\ell$-noZ}}
\newcommand*{\threez}{\mbox{3$\ell$-Z}}

\newcommand*{\vxsttw}{369}
\newcommand*{\vxsttwp}{+100}
\newcommand*{\vxsttwm}{-91}
\newcommand*{\vxsttwstp}{+86}
\newcommand*{\vxsttwstm}{-79}
\newcommand*{\vxsttwsy}{44}
\newcommand*{\vxsttz}{176}
\newcommand*{\vxsttzp}{+58}
\newcommand*{\vxsttzm}{-52}
\newcommand*{\vxsttzstp}{+52}
\newcommand*{\vxsttzstm}{-48}
\newcommand*{\vxsttzsy}{24}

\newcommand*{\pttf}{\ensuremath{p_{\text{T34}}}\xspace}
\newcommand*{\ptcth}{\ensuremath{\pT^{\text{cone30}}}\xspace}
\newcommand*{\etctw}{\ensuremath{\ET^{\text{cone20}}}\xspace}
\renewcommand*{\ptcth}{\ensuremath{\pT^{\Delta R < 0.3}}\xspace}
\renewcommand*{\etctw}{\ensuremath{\ET^{\Delta R < 0.2}}\xspace}
\newcommand*{\mll}{\ensuremath{m_{\ell\ell}}\xspace}
\newcommand*{\lumi}{\ensuremath{20.3~\ifb}\xspace}

\newcommand\mjjvnonbtag{\ensuremath{m_{uu}^{p_{\mathrm{T,ord}}}}}
\newcommand\mjjmindr{\ensuremath{m_{jj}^{\mathrm {min} ~\Delta R}}}
\newcommand\mlepmindrbjetmax{\ensuremath{\mathrm{max} ~m_{\ell b}^{\mathrm {min} ~\Delta R}}}
\newcommand*{\ptjetthree}{\ensuremath{\pT^{\text{jet}3}}\xspace}
\newcommand*{\ptjetfour}{\ensuremath{\pT^{\text{jet}4}}\xspace}
\newcommand*{\drjjav}{\ensuremath{\Delta R^{jj}_{\text{ave}}}\xspace}
\newcommand\numvectorlikethirty{\ensuremath{N_{\mathrm{jet}}^{|m_{jj}-m_{V}| < 30}}}
\newcommand\numjetforty{\ensuremath{N^{\mathrm{jet}}_{40}}}
\newcommand\mbbmaxpt{\ensuremath{m_{bb}^{\mathrm{max} ~p_\mathrm{T}}}}
\newcommand\drleponetwo{\ensuremath{\Delta R_{\ell_1 \ell_2}}}
\newcommand\mbjmaxpt{\ensuremath{m_{bj}^{\mathrm{max} ~p_\mathrm{T}}}}
\newcommand\njets{\ensuremath{N_{\mathrm{jets}}}}
\newcommand*{\hthad}{\ensuremath{\HT^{\text{had}}}\xspace}

\newcommand{\TabLeftBracket}{\begin{tabular}{@{}r@{}}\ldelim\{{2}{2ex} \\ 
\\\end{tabular}}
\newcommand{\TabRightBracket}{\begin{tabular}{@{}l@{}}\rdelim\}{2}{2ex} \\
\\\end{tabular}}

\AtlasTitle{%
Measurement of the $t\bar{t}W$ and $t\bar{t}Z$ production cross sections in $pp$
collisions at $\sqrt{s} = 8 \mathrm{\ Te\kern -0.1em V}$ with the ATLAS detector}

\author{The ATLAS Collaboration}
\PreprintIdNumber{CERN-PH-EP-2015-208}

\AtlasJournal{JHEP}

\AtlasAbstract{%
The production cross sections of top-quark pairs in association with massive
vector bosons have been measured using data from $pp$ collisions at $\sqrt{s} =
8 \mathrm{\ Te\kern -0.1em V}$. The dataset corresponds to an integrated
luminosity of $20.3\ \mathrm{fb}^{-1}$ collected by the ATLAS detector in 2012
at the LHC.  Final states with two, three or four leptons are considered.  A
fit to the data considering the $t\bar{t}W$ and $t\bar{t}Z$ processes
simultaneously yields a significance of $5.0\sigma$ ($4.2\sigma$) over the
background-only hypothesis for $t\bar{t}W$ ($t\bar{t}Z$) production.  The
measured cross sections are \mbox{$\sigma_{t\bar{t}W} = 369^{+100}_{-91}$ fb}
and \mbox{$\sigma_{t\bar{t}Z} = 176^{+58}_{-52}$ fb}.  The background-only
hypothesis with neither $t\bar{t}W$ nor $t\bar{t}Z$ production is excluded at
$7.1\sigma$.  All measurements are consistent with next-to-leading-order
calculations for the $t\bar{t}W$ and $t\bar{t}Z$ processes.}

\hypersetup{pdftitle={ATLAS},pdfauthor={The ATLAS Collaboration}}

\begin{document}
\maketitle
\tableofcontents

\section{Introduction}

The top quark is the heaviest known elementary particle, and its large coupling
to the Higgs boson suggests that it might be closely connected to electroweak
(EW) symmetry breaking. Despite the fact that the top quark was discovered two
decades ago~\cite{Abe:1995hr, Abachi:1995iq} some of its properties, in
particular, its coupling to the $Z$ boson, have never been directly measured.
Several extensions of the Standard Model, such as
technicolour~\cite{Chivukula:1992ap, Chivukula:1994mn, Hagiwara:1995jx,
Mahanta:1996ng, Mahanta:1996qe} or other scenarios with a strongly coupled
Higgs sector~\cite{Perelstein:2005ka} modify the top quark couplings.

\begin{figure}[htbp]
\centering
\subfigure[]{\raisebox{3ex}{\includegraphics[width=0.35\textwidth]{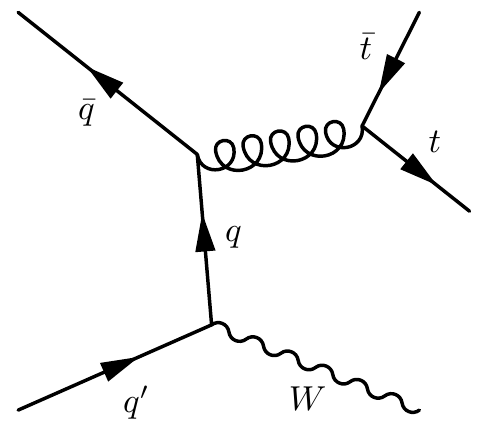}}}
\hspace{2cm}
\subfigure[]{\includegraphics[width=0.37\textwidth]{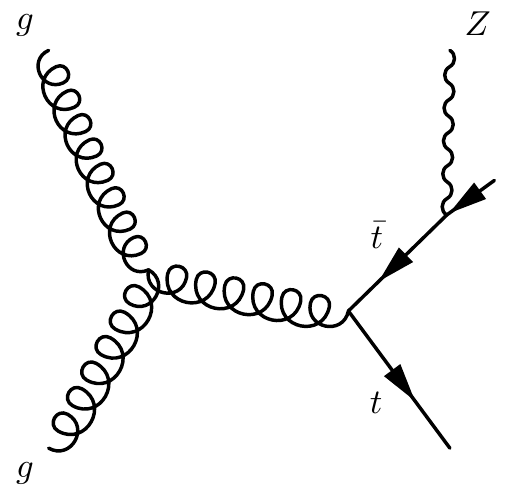}}
\caption{Example leading-order Feynman diagrams for (a) \ttW and (b) \ttZ production.}
\label{fig:FeynmanDiagrams} 
\end{figure}

With the centre-of-mass energy and integrated luminosity of the collected data
samples at the Large Hadron Collider (LHC), the processes in which the
electroweak Standard Model bosons ($\gamma, Z, W$ and $H$) are produced in
association with top quarks become experimentally accessible. Measurements of
the \ttZ, $t\bar{t}\gamma$ and \ttH processes provide a means of
directly determining top quark couplings to bosons~\cite{Baur:2004uw,
Baur:2005wi, Baur:2006ck}, while the \ttW process is a Standard Model (SM)
source of same-sign dilepton events, which are a signature of many models of
physics beyond the SM. Example leading-order Feynman diagrams for \ttW and \ttZ
production at the LHC are shown in Figure~\ref{fig:FeynmanDiagrams}. Previous
searches for \ttW and \ttZ production at the LHC have been carried out by the
CMS collaboration at $\sqrt{s}=7$ TeV and $\sqrt{s} = 8$ TeV
~\cite{Chatrchyan:2013qca, Khachatryan:2014ewa}.

This paper presents measurements of the \ttW and \ttZ cross sections based on
an analysis of \lumi of proton--proton ($pp$) collision data at $\sqrt{s} = 8$
TeV collected by the ATLAS detector. Depending on the decays of the top quarks,
$W$ and $Z$ bosons, between zero and four prompt, isolated leptons\footnote{In
this note, lepton is used to denote electron or muon, including those coming 
from leptonic tau decays.} may be produced.
Channels with two (both with same-sign and opposite-sign charge), three, and
four leptons are considered in this analysis. The opposite-sign (OS) dilepton,
\TL\ and \FL\ channels are mostly sensitive to \ttZ production, while the
same-sign (SS) dilepton channel targets \ttW production.
Table~\ref{tab:intro-channels} lists the analysis channels and the targeted
decay modes of the \ttW and \ttZ processes.  Each channel is divided into
multiple analysis regions in order to enhance the sensitivity to the signal. A
simultaneous fit is performed to all signal regions and selected control
regions in the four channels to extract cross sections for \ttW and \ttZ
production. 

\begin{table}[htbp]
\centering \renewcommand{\arraystretch}{1.2}
\begin{tabular}{|ccccc|}
\hline
Process & \ttbar decay & Boson decay & Channel & $Z \to \ell^+ \ell^-$ \\
\hline
\hline
\multirow{4}{*}{$\ttW^{\pm}$} &  $(\ell^{\pm}\nu b) (q\bar{q} b)$ &
$\ell^{\mp}\nu$ & OS dilepton & no \\
&  $(\ell^{\pm}\nu b) (\ell^{\mp}\nu b)$ & $q\bar{q}$ & OS dilepton & no \\
&  $(\ell^{\pm}\nu b) (q\bar{q} b) $ & $\ell^{\pm}\nu$ & SS dilepton & no\\
& $ (\ell^{\pm}\nu b) (\ell^{\mp}\nu b)$ & $\ell^{\pm}\nu$ & Trilepton & no \\
\hline
\multirow{4}{*}{\ttZ} & $(\ell^{\pm}\nu b) (\ell^{\mp} \nu b)$ & $ q\bar{q}$
& OS dilepton & no   \\
& $(q\bar{q} b) (q\bar{q} b) $ & $\ell^{+}\ell^{-}$ & OS dilepton & yes   \\
& $(\ell^{\pm}\nu b) (q\bar{q} b)$ & $ \ell^{+}\ell^{-}$ & Trilepton & yes \\
& $(\ell^{\pm}\nu b) (\ell^{\mp} \nu b)$ & $ \ell^{+}\ell^{-}$ & Tetralepton &
yes \\
\hline
\end{tabular}
\caption{\label{tab:intro-channels} List of \ttW and \ttZ decay modes and
analysis channels targeting them. The last column indicates whether a final
state lepton pair is expected from a $Z$ boson decay.}
\end{table}

\section{The ATLAS detector}
\label{sec:atlas}

The ATLAS detector~\cite{PERF-2007-01} consists of four main subsystems: an
inner tracking system, electromagnetic (EM) and hadronic calorimeters, and a
muon spectrometer.  The inner detector provides tracking information from pixel
and silicon microstrip detectors in the pseudorapidity\footnote{ATLAS uses a
right-handed coordinate system with its origin at the nominal interaction point
(IP) in the centre of the detector and the $z$-axis coinciding with the axis of
the beam pipe.  The $x$-axis points from the IP to the centre of the LHC ring,
and the $y$-axis points upward.  Cylindrical coordinates ($r$,$\phi$) are used
in the transverse plane, $\phi$ being the azimuthal angle around the beam pipe.
The pseudorapidity is defined in terms of the polar angle $\theta$ as $\eta = -
\ln \tan(\theta/2)$, and the distance between two objects in $\eta-\phi$ space
is measured in terms of $\Delta R \equiv \sqrt{(\Delta \eta)^2 + (\Delta
\phi)^2}$.} range $|\eta|<2.5$ and from a transition radiation tracker (TRT)
covering $|\eta|<2.0$, all immersed in a \SI{2}{T} magnetic field provided by a
superconducting solenoid.  The EM sampling calorimeter uses lead and liquid
argon (LAr) and is divided into barrel ($|\eta|<1.475$) and endcap
($1.375<|\eta|<3.2$) regions.  Hadron calorimetry is provided by a
steel/scintillator-tile calorimeter, segmented into three barrel structures in
the range $|\eta|<1.7$, and two copper/LAr hadronic endcap calorimeters that
cover the region $1.5<|\eta|<3.2$. The solid angle coverage is completed with
forward copper/LAr and tungsten/LAr calorimeter modules, optimised for EM and
hadronic measurements respectively, and covering the region $3.1<|\eta|<4.9$.
The muon spectrometer measures the deflection of muon tracks in the range
$|\eta|<2.7$ using multiple layers of high-precision tracking chambers located
in toroidal magnetic fields of approximately \SI{0.5}{T} and \SI{1}{T} in the
central and endcap regions of ATLAS, respectively. The muon spectrometer is
also instrumented with separate trigger chambers covering $|\eta|<2.4$.

\section{Simulated event samples} 
\label{sec:samples} 

Monte Carlo (MC) samples are used to optimise the event selection and the
choice of signal regions, and to model all signal and certain background
processes.  In the following, the simulation of signal and background events is
described in detail.  For all MC samples, the top quark mass is taken to be
$m_t = \SI{172.5}{\gev}$, and the Higgs boson mass is set to \SI{125}{\gev}. 

The $\ttV (V=W, Z)$ process is simulated using the \textsc{Madgraph5}
leading-order (LO) generator~\cite{Madgraph} with up to one additional parton,
using the \textsc{CTEQ6L1}~\cite{cteq6} parton distribution function (PDF) set.
\textsc{Pythia} 6.425~\cite{PythiaManual} with the AUET2B underlying-event set
of tunable parameters (tune)~\cite{ATL-PHYS-PUB-2011-009} is used to simulate
showering and hadronisation.  The \ttV samples are normalised to the inclusive
next-to-leading-order (NLO) cross-section predictions, using
\textsc{Madgraph5\_aMC@NLO}~\cite{Alwall:2014hca}, including the off-shell
$\ttZ/\gamma^{\ast}$ contribution and interference. An invariant mass of at
least 5 GeV is required for any opposite-sign, same-flavour pair of leptons
appearing in the matrix element. The obtained cross sections are $\sigma_{\ttW}
= 232 \pm \SI{32}{fb}$ and $\sigma_{\ttZ} = 215 \pm \SI{30}{fb}$, compatible
with other NLO QCD calculations~\cite{Campbell:2012dh, Garzelli:2012bn}. The
quoted uncertainties include renormalisation and factorisation scale and PDF
uncertainties, including $\alpha_\mathrm{S}$ variations.

The \textsc{Alpgen v2.14}~\cite{Alpgen} LO generator and the \textsc{CTEQ6L1}
PDF set are used to simulate $W/Z$ production.  Parton showers and
hadronisation are modelled with \textsc{Pythia} 6.425. The $W/Z$ samples are
generated with up to five additional light partons, separately for $W/Z$,
$W/Z$+$\bbbar$, $W/Z$+$\ccbar$ and $Wc$, and normalised to the respective
inclusive next-to-next-to-leading-order (NNLO) theoretical cross
sections~\cite{Melnikov:2006kv}.  To avoid double-counting of partonic
configurations generated by both the matrix-element calculation and the
parton-shower evolution, a parton--jet matching scheme (MLM
matching)~\cite{MLM} is employed. The overlap between $W/Z$+$Q\bar{Q}$
($Q=b,c$) events generated from the matrix-element calculation and those
generated from parton-shower evolution in the $W/Z$+light-jet samples is
avoided via an algorithm based on the distance in $\eta$--$\phi$ space between
the heavy quarks: if $\Delta R(Q,\bar{Q})>0.4$, the matrix-element prediction
is used, otherwise the parton-shower prediction is used. 

Diboson samples are generated using the \textsc{Sherpa} 1.4.1~\cite{Sherpa}
generator with the \textsc{CT10} PDF set~\cite{Lai:2010vv}, with massive $b$-
and $c$-quarks and with up to three additional partons in the LO matrix
element.  Samples are normalised to their NLO QCD theoretical cross
sections~\cite{Campbell:1999ah}. Alternative models for the diboson background
are provided by the \mbox{\textsc{POWHEG-BOX} 2.0} \cite{Powheg, Powbox1,
Powbox2} generator, which implements the NLO matrix elements, interfaced with
\textsc{Pythia} 6.425 or \textsc{Pythia} 8.1~\cite{PythiaManual8}.

Simulated \ttbar and single-top-quark backgrounds corresponding to the
$t$-channel, $Wt$ and $s$-channel production mechanisms are generated using the
\textsc{POWHEG-BOX} generator, with the \textsc{CT10} PDF set. All samples are
interfaced with \textsc{Pythia} 6.425 with the \textsc{CTEQ6L1} PDF set and the
Perugia2011C~\cite{skands:2010ak} underlying event tune. Overlaps between the
\ttbar and $Wt$ final states are removed through the diagram removal
scheme~\cite{Frixione:2005vw}.  The \ttbar sample is normalised to the
Top++2.0~\cite{ref:xs6} theoretical calculation performed at NNLO in QCD that
includes resummation of next-to-next-to-leading logarithmic soft gluon
terms~\cite{ref:xs1,ref:xs2,ref:xs3,ref:xs4,ref:xs5}.  The single-top-quark
samples are normalised to the approximate NNLO theoretical cross
sections~\cite{Kidonakis:2011wy, Kidonakis:2010tc, Kidonakis:2010ux} calculated
using the \textsc{MSTW2008} NNLO PDF set~\cite{Martin:2009iq,Martin:2009bu}. 

The production of a single top quark in association with a $Z$ boson through
the $t$- and $s$-channels, of the $WtZ$ process, and of a top quark pair in
association with a $W$ boson pair (\ttbar$WW$) are simulated with
\textsc{Madgraph5} LO and the \textsc{CTEQ6L1} PDF set. \textsc{Madgraph} is
interfaced with \textsc{Pythia} 6.425 using the AUET2B tune and the
\textsc{CTEQ6L1} PDF set. The relevant samples are normalised to the NLO
theoretical predictions calculated with
\textsc{Madgraph5\_aMC@NLO}~\cite{Alwall:2014hca}.  The production of three
vector bosons that decay to three or four leptons is also simulated with
\textsc{Madgraph5} and \textsc{Pythia} 6.425. The LO cross section obtained
from the generator is used to normalise the samples.  The production of two $W$
bosons with the same charge is modelled using the \textsc{Sherpa} generator,
including diagrams of order $\alpha^4_\mathrm{EW}$ and $\alpha^2_\mathrm{EW}
\alpha^2_\mathrm{S}$. The LO cross section obtained from the generator is
used to normalise the samples.  The four-top-quark process ($\ttbar \ttbar$) is
simulated with \textsc{Madgraph5} interfaced with \textsc{Pythia} 8.

Associated \ttH production is simulated using NLO matrix elements obtained from
the \textsc{HELAC-Oneloop} package~\cite{Helac}.  The \textsc{POWHEG-BOX}
program served as an interface for shower MC programs.  Samples were produced
using the \textsc{CT10NLO} PDF set and showered with \textsc{Pythia} 8.1 with
the \textsc{CTEQ6L1} PDF and the AU2 underlying-event tune~\cite{ATLASUETune}.
The \ttH cross section and Higgs boson decay branching fractions are taken from
the theoretical calculations collected in Ref.~\cite{Dittmaier:2011ti}.  The
process $gg\to H\to 4 \ell$ is modelled using the \textsc{POWHEG-BOX}
interfaced with \textsc{Pythia 8.1}. $WH$ and $ZH$ production are modelled
using \textsc{Pythia 8.1}. The samples are normalised to the NNLO QCD cross
sections with NLO electroweak corrections~\cite{Dittmaier:2011ti}.

All simulated samples produced with \textsc{Pythia} 
use \textsc{Photos 2.15}~\cite{Golonka:2005pn} to simulate photon radiation
and \textsc{Tauola 1.20}~\cite{Jadach:1990mz} to simulate \mbox{$\tau$ decays}.
Events from minimum-bias interactions from the same bunch crossing as the
hard-scattering process and in neighbouring bunch crossings, known as
pile-up, are simulated with the \textsc{Pythia} 8.1 generator with the
\textsc{MSTW2008} LO PDF set  and the AUET2~\cite{ATLASUETune1} tune. These are 
superimposed on the simulated hard-scatter events in a manner which reproduces  
the luminosity profile of the recorded data. 

All samples are processed through a simulation of the detector geometry and
response~\cite{SOFT-2010-01} either using \textsc{Geant4}~\cite{Geant}, or
\textsc{Geant4} with a fast simulation of the calorimeter
response~\cite{ATLASFastSim}.  All samples are processed by the same
reconstruction software as the data. Simulated events are corrected so that the
object identification, reconstruction and trigger efficiencies, energy scales
and energy resolutions match those determined from data control samples.

\section{Object reconstruction}
\label{sec:ObjectReconstruction}

The final states of interest in this analysis contain electrons, muons, jets,
$b$-jets and missing transverse momentum. 

Electron candidates~\cite{PERF-2013-03} are reconstructed from energy deposits
(clusters) in the EM calorimeter that are associated with reconstructed tracks
in the inner detector.  The electrons are required to have
$|\eta_\mathrm{cluster}| < 2.47$, where $\eta_\mathrm{cluster}$ is the
pseudorapidity of the calorimeter energy deposit associated with the electron
candidate.  Candidates in the EM calorimeter barrel/endcap transition region
$1.37 < |\eta_\mathrm{cluster}| < 1.52$ are excluded. The electron
identification relies on a likelihood-based
selection~\cite{ATLAS-CONF-2014-032}.

To reduce the background from misidentified or non-prompt (labelled as ``fake''
throughout this paper) electrons, i.e.~from decays of hadrons (including heavy
flavour), electron candidates are required to be isolated. In the opposite-sign
dilepton and tetralepton channels, in which such background is small, the
electron isolation is defined using only tracking information.  In the \OSLC,
the ratio of \ptcth, the sum of track transverse momenta in a cone of size
$\Delta R = 0.3$ around the electron track, excluding the electron track
itself, to the transverse momentum ($\pT^e$) of the electron is required to be
less than 0.12.  In the \FLC, the requirement is loosened to  $\ptcth/\pT^e <
0.18$. In the \TL\ and \SSL\ channels, in which the background with fake
leptons is more prominent, additional requirements are imposed on the electron
isolation. For electrons with $\pt^e< \SI{50}{\gev}$, both the ratio of the
additional calorimeter energy within a cone of $\Delta R = 0.2$ around the
electron (\etctw) to the $\pT^e$ of the electron, and $\ptcth/\pT^e$ are
required to be less than $0.12$.  For $\pt^e \ge \SI{50}{\gev}$, both $\etctw$
and $\ptcth$ are required to be less than \SI{6}{\gev}. 

Muon candidates are reconstructed from track segments in the various layers of
the muon spectro\-meter, and matched with tracks identified in the inner
detector~\cite{PERF-2014-05}. The final muon candidates are refitted using the
complete track information from both detector systems, and are required to have
$|\eta|<2.5$.  Additionally, muons are required to be separated by $\Delta R >
0.4 $ from any jet and to satisfy a $\pt$-dependent track-based isolation
requirement~\cite{Rehermann:2010vq} that has good performance under high
pile-up conditions. This requires that the scalar sum of the track transverse
momenta in a cone of variable size $\Delta R = (\SI{10}{\gev}/\pt^\mu)$ around
the muon (excluding the muon track itself) must be less than $0.05\pt^\mu$.

For both the electrons and muons, the track longitudinal impact parameter with
respect to the primary vertex,\footnote{A primary vertex candidate is defined
as a vertex with at least five associated tracks, consistent with the beam
collision region.  If more than one such vertex is found, the vertex candidate
with the largest sum of squared transverse momenta of its associated tracks is
taken as the primary vertex.} $z_{0}$, is required to be less than \SI{2}{mm}.
In the \SSLC, in which backgrounds from fake leptons are dominant, it is also
required to satisfy $|z_0 \sin\theta|<0.4$ mm, and the significance of the
transverse impact parameter $d_0$ is required to satisfy $|d_0/\sigma(d_0)|<3$,
where $\sigma(d_0)$ is the uncertainty on $d_0$.  

Jets are reconstructed with the anti-$k_t$ algorithm~\cite{Cacciari:2008gp,
Cacciari:2005hq, Fastjet} with radius parameter $R=0.4$ from calibrated
topological clusters~\cite{PERF-2007-01} built from energy deposits in the
calorimeters.  Prior to jet finding, a local cluster calibration
scheme~\cite{Cojocaru:2004jk, ATL-LARG-PUB-2008-002} is applied to correct the
topological cluster energies for the effects of non-compensating calorimeter
response, dead material and out-of-cluster leakage.  The jets are calibrated to
restore the jet energy scale to that of jets reconstructed from stable
simulated particles, using energy- and $\eta$-dependent calibration factors
derived from simulations.  Additional corrections to account for residual
differences between simulation and data are applied~\cite{PERF-2012-01}. After
calibration, jets are required to have $\pt > \SI{25}{\gev}$ and $|\eta| <
2.5$.  To avoid selecting jets from pile-up interactions, an additional
requirement, referred to as the jet vertex fraction criterion (JVF), is imposed
on jets with $\pt < \SI{50}{\gev}$ and $|\eta|<2.4$.  It requires that at least
50\% of the scalar sum of the transverse momenta of tracks with $\pt >
\SI{1}{\gev}$, associated with a jet, comes from tracks compatible with
originating from the primary vertex.  During jet reconstruction, no distinction
is made between identified electrons and jet energy deposits.  Therefore, if
any of the jets lie $\Delta R<0.2$ from an electron, the closest jet is
discarded in order to avoid double counting of electrons as jets.  After this
overlap removal, electrons and muons which lie $\Delta R<0.4$ from any
remaining jet are removed. 

Jets containing $b$-hadrons are tagged by an algorithm
(MV1)~\cite{ATLAS-CONF-2011-102} that uses multivariate techniques to compute
weights by combining information from the impact parameters of displaced tracks
as well as topological properties of secondary and tertiary decay vertices
reconstructed within the jet.  Larger weights indicate that a jet is more
likely to contain $b$-hadrons.  The working point used for this measurement
corresponds to 70\% efficiency to tag a $b$-quark jet, as determined for
$b$-jets with $\pt > \SI{20}{\gev}$ and $|\eta|<2.5$ in simulated \ttbar
events.  The rejection factors for light-jets and $c$-quark jets are
approximately 130 and 5, respectively.  The efficiency of $b$-tagging in
simulation is corrected to that in data using a \ttbar based
calibration~\cite{ATLAS-CONF-2014-004}.  

The missing transverse momentum $\mathbf{p}^\mathrm{miss}_\mathrm{T}$, with magnitude
\met, is reconstructed~\cite{PERF-2011-07} as the negative sum of transverse
momenta of all electrons, muons, jets and calibrated calorimeter energy
clusters not associated with any of these objects.

\section{Event selection and background estimation}
\label{sec:EventSelection}

The measurements presented here are based on data collected by the ATLAS
experiment in $pp$ collisions at $\sqrt{s}=8\tev$ in 2012. The corresponding
integrated luminosity is \lumi.  Only events collected using a single-electron
or single-muon trigger under stable beam conditions, that satisfy the standard
data quality criteria, are accepted.  The trigger \pT thresholds are 24 or
\SI{60}{\gev} for electrons and 24 or \SI{36}{\gev} for muons: the triggers
with the lower \pT thresholds include isolation requirements on the candidate
lepton, resulting in inefficiencies at high \pT that are recovered by the
triggers with higher \pT thresholds. Events are required to have at least one
reconstructed primary vertex.  In all selections considered, at least one
lepton with $\pT>\SI{25}{\gev}$ is required to match \mbox{($\Delta R<0.15$)} a
lepton with the same flavour, reconstructed by the trigger algorithm. 

Four channels are defined based on the number and charges of the reconstructed
leptons, which are sorted according to their transverse momentum in decreasing
order. For the \OSLC, two leptons with opposite charge and $\pT>\SI{15}{\gev}$
are required. In the \SSLC, events are required to contain two same-sign
leptons with $\pT>\SI{25}{\gev}$. In both dilepton channels, events containing
additional leptons with $\pT>\SI{15}{\gev}$ are rejected. For the \TLC, events
are required to contain three leptons with $\pT>\SI{15}{\gev}$. For the \FLC,
exactly four leptons with $\pT>\SI{7}{\gev}$ are required. Events satisfying
both the trilepton and \FLC\ selections are attributed to the \TLC\ and removed
from the \FLC. The dilepton channels are not explicitly required to be
orthogonal with the \FLC, but the overlap is found to be negligibly small in
simulated samples and non-existent in data.

Background events containing well-identified prompt leptons are modelled by
simulation.  The normalisations used for the backgrounds in this category are
taken from data control regions if the resulting normalisation uncertainty is
lower than that from the theoretical prediction.  The yields in the data
control regions are extrapolated to the signal regions using the simulation.
Background sources involving one or more incorrectly identified lepton, e.g.
instrumental backgrounds, are modelled using data events from control regions,
except in the \OSLC, where this background is very small. 

The following sections describe additional selection requirements and the
background evaluation in each of the four channels. 

\subsection{Opposite-sign dilepton channel}
\label{sec:OSLC}

In the \OSLC, events are required to have at least three jets, one or two of
which are $b$-tagged.  Two orthogonal selections are defined to separate \ttW
and \ttZ final states.  The first (\rone) selects different-flavour lepton (DF)
events with the scalar sum of the \pT of leptons and jets, $\HT$, above
${130}{\gev}$,  and same-flavour lepton (SF) events that are not compatible
with \Zboson boson or low-mass resonance production, by requiring $|\mll -
m_Z|>\SI{10}{\gev}$, $\mll>\SI{15}{\gev}$ and \mbox{$\met>\SI{40}{\gev}$}.  The
\rtwo\ selection contains SF events within the mass window \mbox{$|\mll -
m_Z|<\SI{10}{\gev}$}.  In both selections an additional requirement on the
average distance between two jets, calculated using all possible jet pairs in
the event, $\drjjav>0.75$, is applied to remove the low-dijet-mass region where
the \textsc{Alpgen}+\textsc{Pythia} simulation does not provide a good
description of the $Z$ boson background~\cite{STDM-2012-04}. 

For \rone\ events, the \ttV signal contribution originates mainly from the
\ttbar dilepton final state accompanied by a hadronic $W/Z$ boson decay and
from the \ttbar single-lepton final state with a leptonic $W$ boson decay.  For
\rtwo\ events, the contribution of \ttW production is negligible while the \ttZ
contribution comes from the fully hadronic \ttbar final state with a leptonic
$Z$ boson decay.

After event selection the dominant backgrounds are  \ttbar and $Z$ production
in \rone\ and \rtwo\, respectively, and the extraction of the signal relies on
discriminating it from these backgrounds, based on well-modelled event
kinematics.  To improve the modelling of the \ttbar background, the simulated
\ttbar events are reweighted to account for the observed differences in the top
quark \pT and the \ttbar system \pT between data and
\textsc{Powheg}+\textsc{Pythia} simulation in measurements of differential
cross sections at $\sqrt{s}= \SI{7}{\tev}$~\cite{topdiff_7TEV}.  To improve $Z$
background modelling, the simulation is reweighted to account for the
difference in the $Z$ \pT spectrum between data and
simulation~\cite{STDM-2012-04}, and the $ZQ\bar{Q}$ ($Q=b,c$) component of the
$Z$ background is adjusted to match data in a $ZQ\bar{Q}$-dominated control
region with at least one $b$-tagged jet.  Small background contributions arise
from single-top-quark $Wt$ channel production, diboson ($WW$, $WZ$, $ZZ$)
processes, the associated production of a Higgs boson and a \ttbar pair, the
associated production of a $WW$ and a \ttbar pair, and the associated
production of a single top quark and a $Z$ boson. All of these backgrounds are
determined from simulation.

In the \rone\ region, $W$ boson, \ttbar (with a single lepton  in the final
state) and $t$- and $s$-channel single-top-quark production processes can
satisfy the selection requirements due to fake leptons.  These backgrounds are
a small fraction of the total estimated background, and their yields are
estimated using simulation and cross-checked with a data-driven technique based
on the selection of a same-sign lepton pair.

Events are categorised according to the number of jets and the number of
$b$-tagged jets.  In the \rone\ selection, events with one or two $b$-tagged
jets are separated into three exclusive regions according to the jet
multiplicity, with three (\OSLCRA), four (\OSLSRA), and five or more (\OSLSRB)
jets.  In the \rtwo\ selection, events with exactly two $b$-tagged jets are
separated into three regions according to the same scheme: \OSLCRB, \OSLCRC\
and \OSLSRC. 

A neural network (NN) discriminant built using the NeuroBayes~\cite{Neurobayes}
package is used to separate the combined \ttW and \ttZ signal from the
background in the signal-rich regions \OSLSRA, \OSLSRB\ and \OSLSRC.  The other
regions considered in the \OSLC\ have lower sensitivity and are used as control
regions; event counting is used in the \OSLCRA\ region, while the scalar sum of
the jet transverse momenta (\hthad) is used as a discriminant in the \OSLCRB\
and \OSLCRC\ regions.  The inclusion of these highly populated control regions,
enriched in \ttbar or $Z$ backgrounds, in the fit used to extract the \ttV
signals, strongly constrains the normalisation uncertainties of these
backgrounds. This in turn improves the background predictions in the
signal-rich regions. The signal and control regions are summarised in
Table~\ref{tab:OSregions}.

\begin{table}[htbp]
\centering \renewcommand{\arraystretch}{1.2}
\begin{tabular}{|ccc|}
\hline
Region & Targeting & Sample fraction [\%] \\
\hline
\hline
\OSLSRA  & \multirow{2}{*}{\ttW and \ttZ} & 0.68 \\
\OSLSRB  & & 1.2\\ \hline
\OSLSRC & $\ttZ$ & 3.3\\
\OSLCRA & $t \bar t$ & 92 \\ \hline
\OSLCRB  & \multirow{2}{*}{$Z$} & 70\\
\OSLCRC  &  & 66 \\
\hline
\end{tabular}
\caption{\label{tab:OSregions} Signal and control regions of the \OSLC,
together with the processes targeted and the expected fraction of the sample
represented by the targeted process.}
\end{table}

The set of variables used as input to the NN discriminant is chosen separately
for each signal region, based on the ranking procedure implemented in the
NeuroBayes package which takes into account the statistical separation power of
the variables and the correlations between them.  All variables used for the NN
training are required to show good agreement between data and background
expectation in the control regions. Seven variables are selected in each signal
region.  The list of selected variables and their ranking is shown in
Table~\ref{tab:NeuralNetVariables}.

\begin{table}[htbp]
\centering \renewcommand{\arraystretch}{1.2}
\begin{tabular*}{\textwidth}{|lp{0.45\textwidth}@{\extracolsep{\fill}}ccc|}
\hline
\multirow{ 2}{*}{Variable} & \multirow{ 2}{*}{Definition}  &
\multicolumn{2}{r}{NN rank} & \\
& &  \OSLSRA\ & \OSLSRB\ & \OSLSRC\ \\ 
\hline
\hline
\mjjvnonbtag & Invariant mass of the two highest \pt untagged jets in events
with exactly two $b$-tags, or of the two highest \pt untagged jets, excluding
the jet with the second highest $b$-tag weight, in events with exactly one
$b$-tag  & 1st & 7th & - \\
$\mathrm{Centrality}_\mathrm{jet}$   & Sum of $\pt$ divided by sum of $E$ for all
jets & 2nd & 1st & 6th \\
$H_1$ & 2nd Fox-Wolfram moment~\cite{Fox:1978vu} & 3rd & 2nd & - \\
\mjjmindr & Invariant mass of the combination of the two jets with the smallest
$\Delta R$ & 4th & 6th & - \\
\mlepmindrbjetmax  & Larger of the invariant masses of the two (lepton,
$b$-tagged jet) pairs, which are built based on the minimum $\Delta R(\ell, b)$
for each lepton & 5th & 5th & - \\
\ptjetthree & Third-jet $\pt$ & 6th & - & - \\
\ptjetfour & Fourth-jet $\pt$ & - & 3rd & - \\
\drjjav & Average $\Delta R$ for all jet pairs & 7th & - & - \\
\numvectorlikethirty  & Number of jet pairs with mass within a \SI{30}{\gev}
window around \SI{85}{\gev} & - & 4th & 2nd \\
\numjetforty & Number of jets with $\pt > 40\GeV$ & - & - & 1st \\
\mbbmaxpt & Invariant mass of the combination of two $b$-tagged jets with the
largest vector sum $\pt$ & - & - & 3rd \\
\drleponetwo & $\Delta R$ between the two leptons & - & - & 4th\\
\mbjmaxpt & Invariant mass of the combination of the two jets with the largest
vector sum $\pt$; one jet must be $b$-tagged & - & - & 5th \\
$H_{1}^{\text{jet}}$ & 2nd Fox-Wolfram moment built from only jets & - & - & 7th\\
\hline
\end{tabular*}
\caption{Definitions and rankings of the variables considered in each of the
regions where a NN is used in the \OSLC. }
\label{tab:NeuralNetVariables}
\end{table}

\begin{figure}[htbp]
\centering
\subfigure[]{\includegraphics[height=0.37\textwidth]{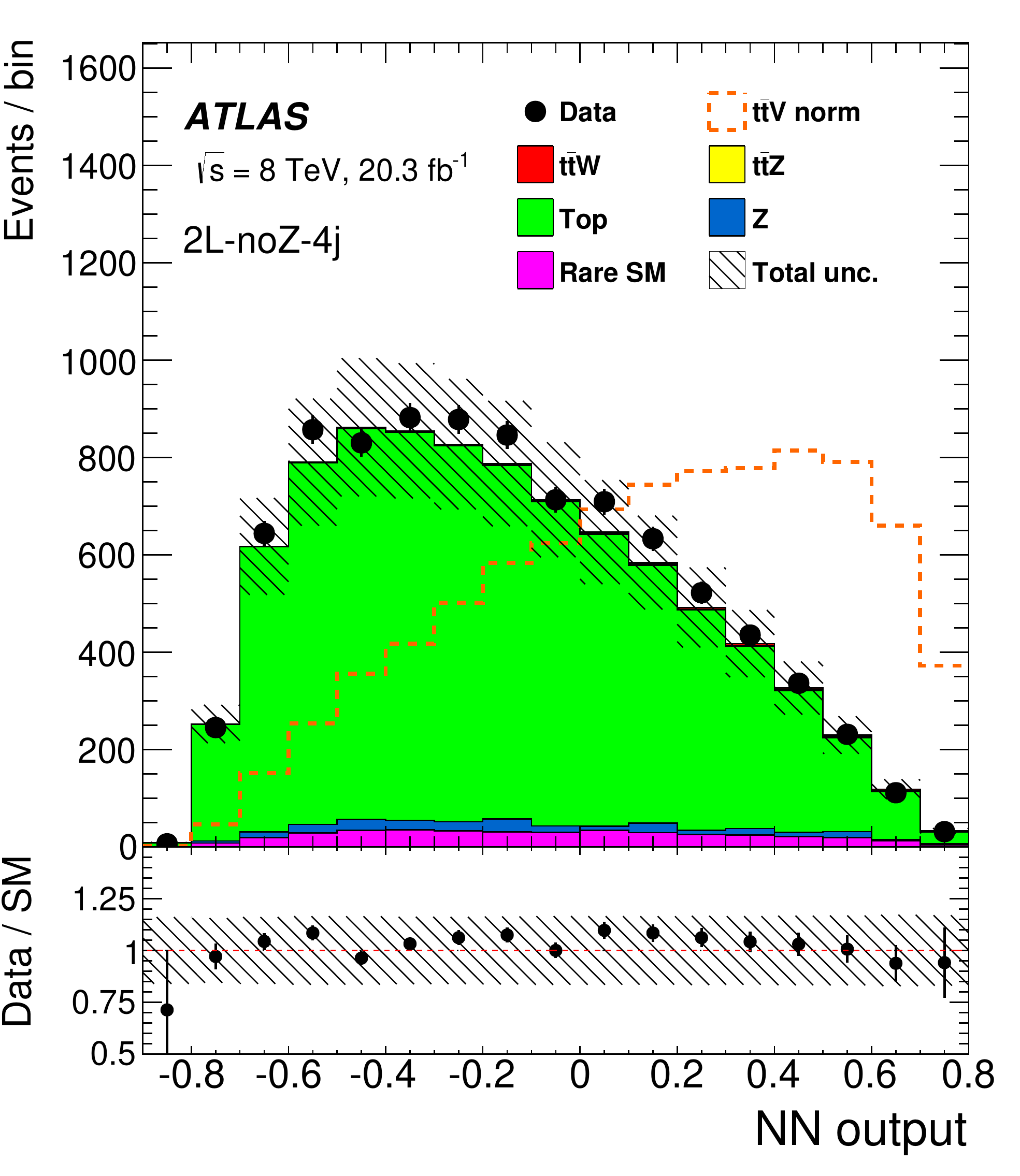}}
\subfigure[]{\includegraphics[height=0.37\textwidth]{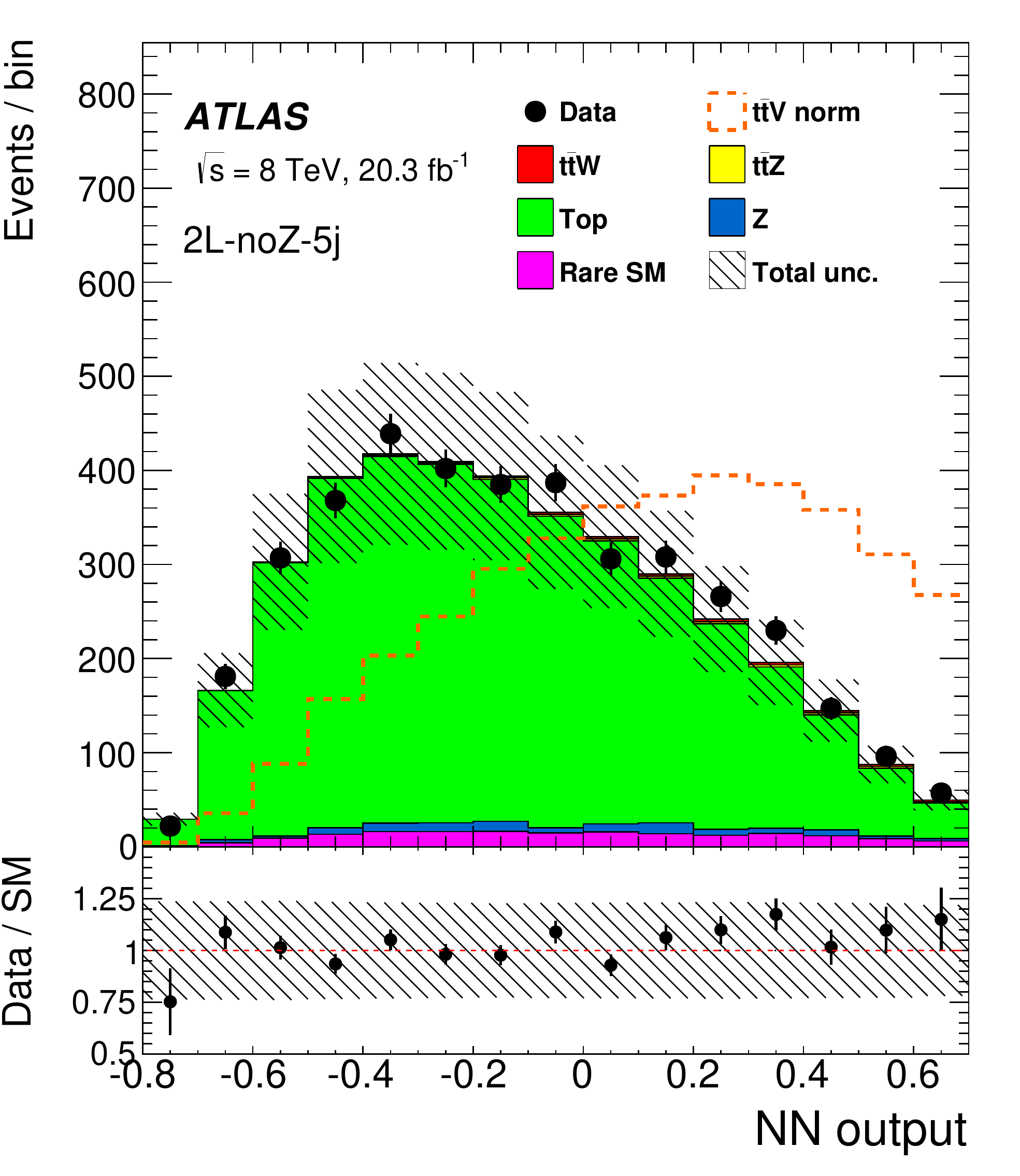}}
\subfigure[]{\includegraphics[height=0.37\textwidth]{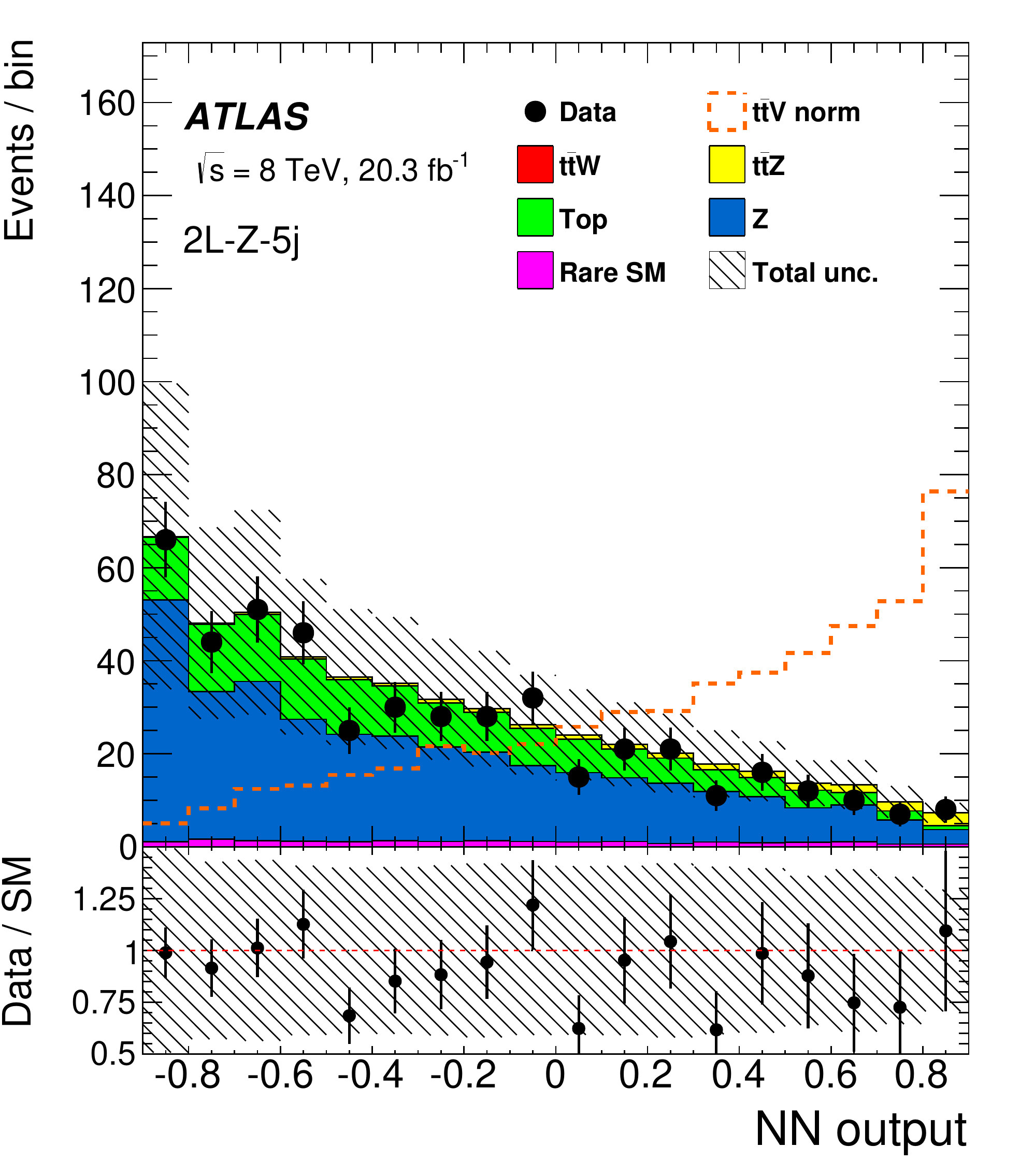}}
\caption{The NN output distributions for the three signal regions in the \OSLC,
before the fit to data. The distributions are shown in the (a) \OSLSRA, (b)
\OSLSRB\ and (c) \OSLSRC\ regions. The orange dashed lines show the \ttV signal
normalised to the background yield.  ``Rare SM'' comprises the diboson,
single-top, $tZ$, $WtZ$, \ttH processes and the fake lepton background.
The hatched area corresponds to the total uncertainty on the predicted yields.
The ``Data/SM'' plots show the ratio of the data to the total Standard Model
expectation.}
\label{fig:OSL}
\end{figure}

\Fig{OSL} illustrates the discrimination between the \ttV signal and background
provided by the NN discriminants.  
Since in the \rone\ region the contributions from both \ttW\ and \ttZ\ production 
are comparable in size and have similar
kinematics, they result in a similar NN discriminant shape and are thus fitted
together. In the \rtwo\ region, the \ttW contribution is negligible, and thus
the NN discriminant shape is driven by the \ttZ signal.   

\begin{figure}[htbp]
\centering
\subfigure[]{\includegraphics[height=0.37\textwidth]{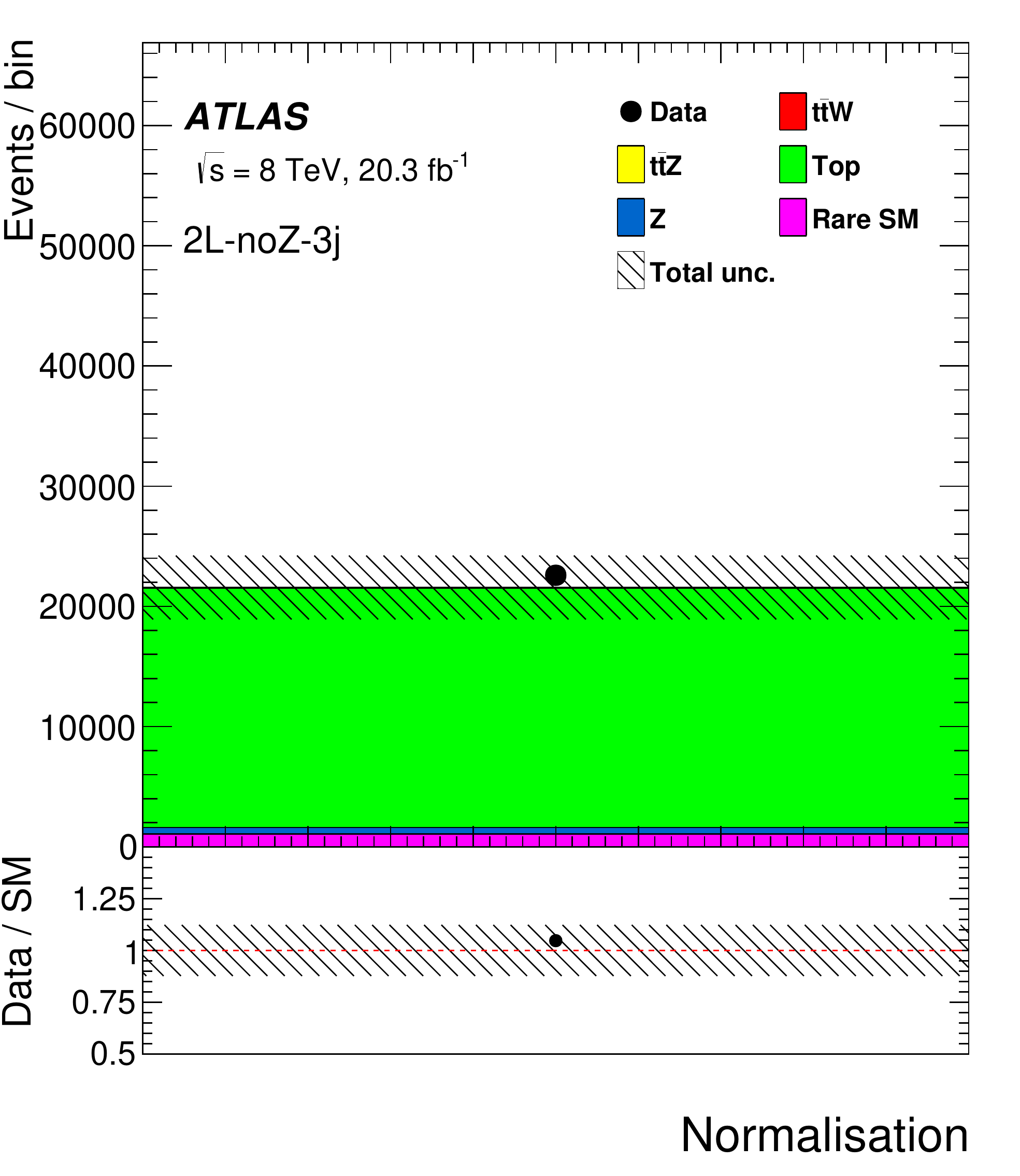}}
\subfigure[]{\includegraphics[height=0.37\textwidth]{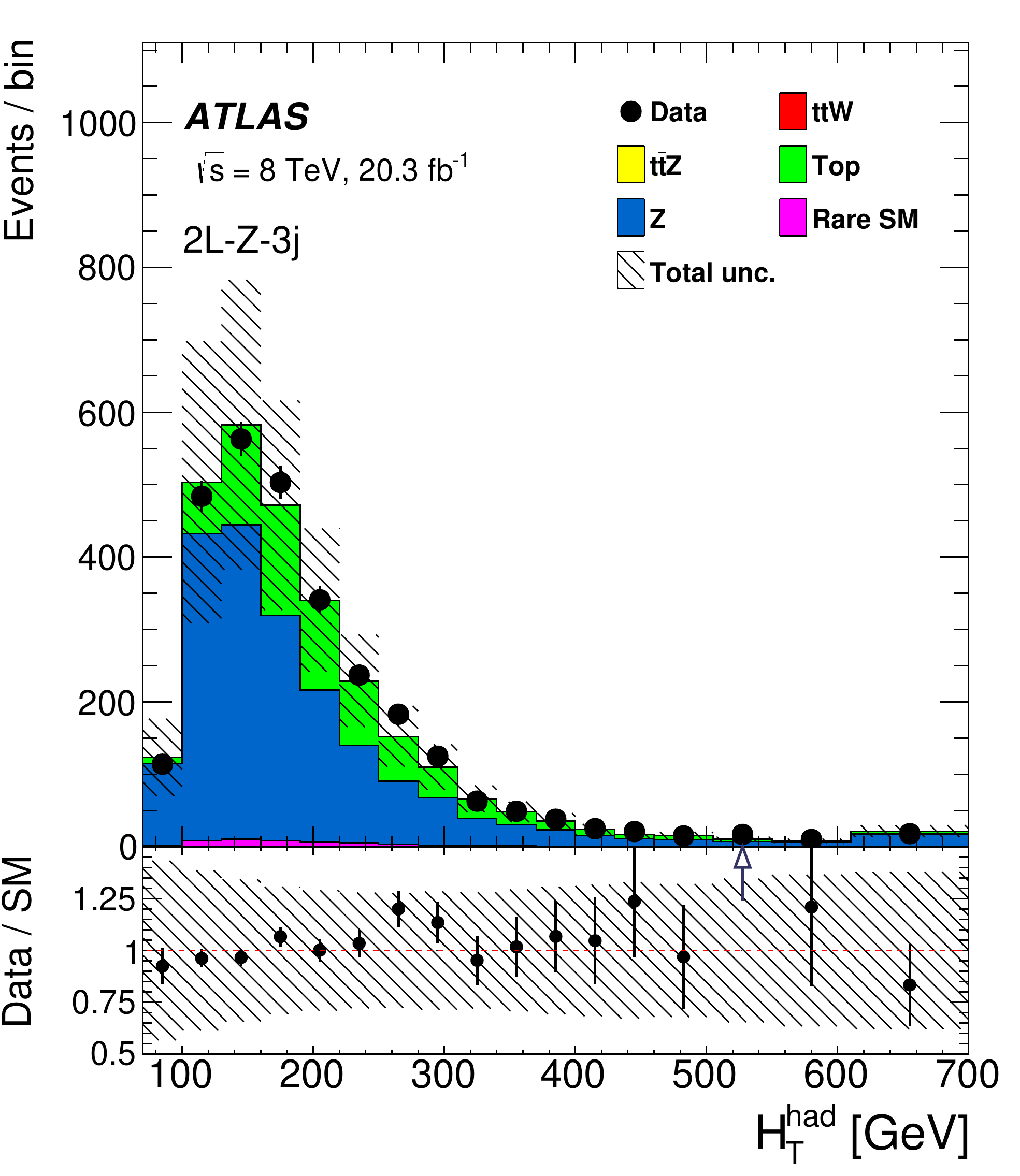}}
\subfigure[]{\includegraphics[height=0.37\textwidth]{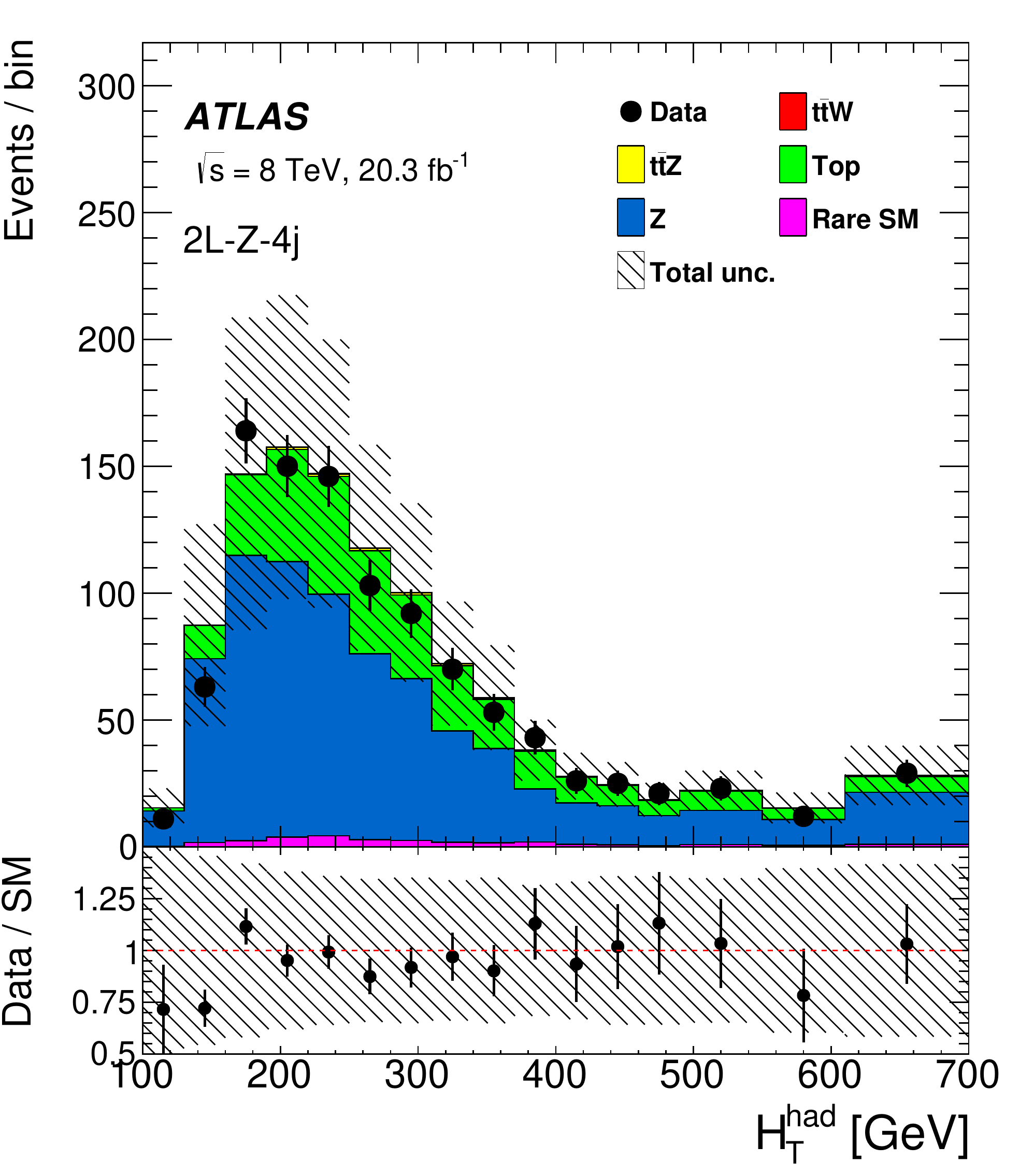}}
\caption{Control region distributions in the \OSLC, before the fit to data. The
distributions are shown in the (a) \OSLCRA, (b) \OSLCRB\ and (c) \OSLCRC\
regions. For the \OSLCRA\ region only the event count is used in the fit. The
hatched area corresponds to the total uncertainty on the predicted yields.
``Rare SM'' comprises the diboson, single-top, $tZ$, $WtZ$, \ttH
processes and the fake lepton background. 
The ``Data/SM'' plots show the
ratio of the data to the total expected Standard Model expectation.}
\label{fig:OSL-CR}
\end{figure}

The expected sample compositions in each of the three signal and three control
regions are summarised in \Tab{expected_yields} along with the number of events
observed in data.  The distributions of discriminants in the control regions
are shown in \Fig{OSL-CR}.  The data and simulation agree within the expected
uncertainties. 

\begin{table}[htbp]
\centering \renewcommand{\arraystretch}{1.2}
\resizebox{\columnwidth}{!}{
\begin{tabular}{%
|c
r@{\,}@{$\pm$}@{\,}l
r@{\,}@{$\pm$}@{\,}l
r@{\,}@{$\pm$}@{\,}l
r@{\,}@{$\pm$}@{\,}l
r@{\,}@{$\pm$}@{\,}l
r@{\,}@{$\pm$}@{\,}l
r|
}
\hline
Region & 
\multicolumn{2}{c}{$t+X$} &
\multicolumn{2}{c}{Bosons} &
\multicolumn{2}{c}{Fake leptons} & 
\multicolumn{2}{c}{Total expected} &
\multicolumn{2}{c}{\ttW} &
\multicolumn{2}{c}{\ttZ} &
Data \\
&
\multicolumn{4}{c}{} & 
\multicolumn{2}{c}{charge misID} &
\multicolumn{2}{c}{background} & 
\multicolumn{4}{c}{} & \\ 
\hline
\hline
\OSLCRA* & 20800 & 2600 & 600 & 200 & 160 & 80 & 21600 & 2700 & 42.0 & 2.8 & 23.2 & 1.5 & 22585\\
\OSLSRA & 8200 & 1400 & 240 & 90 & 80 & 40 & 8600 & 1400 & 36.6 & 1.8 & 22.4 & 1.1 & 8909\\
\OSLSRB & 3700 & 850 & 100 & 40 & 47 & 23 & 3810 & 870 & 24.9 & 2.2 & 22.4 & 2.0 & 3901\\
\OSLCRB* & 800 & 140 & 1960 & 880 & 4.1 & 2.1 & 2760 & 890 & 1.24 & 0.13 & 3.71 & 0.38 & 2806\\
\OSLCRC* & 330 & 70 & 740 & 390 & 2.2 & 1.1 & 1100 & 400 & 1.31 & 0.11 & 7.21 & 0.58 & 1031\\
\OSLSRC & 170 & 40 & 340 & 200 & 1.4 & 0.7 & 510 & 210 & 0.89 & 0.07 & 17.7 & 1.4 & 471\\\hline
\SSLSRA & 0.66 & 0.13 & 0.17 & 0.10 & 8.9 & 2.4 & 9.8 & 2.6 & 2.97 & 0.30 & 0.93 & 0.23 & 16\\
\SSLSRB & 1.9 & 0.35 & 0.39 & 0.28 & 14.1 & 4.5 & 16.4 & 5.1 & 8.67 & 0.76 & 2.16 & 0.51 & 34\\
\SSLSRC & 0.94 & 0.17 & 0.25 & 0.14 & 0.93 & 0.55 & 2.12 & 0.86 & 4.79 & 0.40 & 1.12 & 0.27 & 13\\\hline
\TLCR* & 1.11 & 0.32 & 67 & 16 & 15.2 & 6.0 & 83 & 15 & 0.05 & 0.03 & 1.86 & 0.47 & 86\\
\TLSRA & 1.58 & 0.42 & 3.8 & 1.3 & 2.4 & 1.1 & 7.8 & 1.6 & 0.14 & 0.05 & 7.1 & 1.6 & 8\\
\TLSRB & 1.29 & 0.34 & 0.68 & 0.33 & 0.19 & 0.13 & 2.16 & 0.42 & 0.21 & 0.07 & 2.76 & 0.69 & 3\\
\TLSRC & 1.00 & 0.29 & 0.48 & 0.24 & 0.42 & 0.37 & 1.93 & 0.49 & 0.14 & 0.07 & 6.6 & 1.6 & 11\\
\TLSRD & 1.06 & 0.25 & 0.27 & 0.17 & 1.31 & 0.90 & 2.7 & 0.9 & 3.7 & 0.9 & 1.23 & 0.32 & 6\\\hline
\FLSRA & 0.06 & 0.01 & 0.11 & 0.04 & 0.03 & 0.17 & 0.21 & 0.22 &\multicolumn{2}{c}{-} & 0.28 & 0.01 & 2\\
\FLSRB & 0.22 & 0.03 & 0.05 & 0.03 & 0.13 & 0.22 & 0.39 & 0.27 &\multicolumn{2}{c}{-} & 1.05 & 0.03 & 1\\
\FLSRC & 0.11 & 0.02 & \multicolumn{2}{c}{<0.01} &0.11 & 0.19 & 0.22 & 0.21 &\multicolumn{2}{c}{-} & 0.64 & 0.02 & 1\\
\FLCR* & 0.01 & 0.00 & 134.2 & 1.2 & 0.27 & 0.18 & 134.5 & 1.3 &\multicolumn{2}{c}{-} & 0.07 & 0.01 & 158\\
\FLSRD & 0.16 & 0.02 & 0.29 & 0.06 & 0.14 & 0.19 & 0.61 & 0.27 &\multicolumn{2}{c}{-} & 0.91 & 0.02 & 2\\
\FLSRE & 0.08 & 0.01 & 0.09 & 0.03 & 0.04 & 0.18 & 0.21 & 0.23 &\multicolumn{2}{c}{-} & 0.64 & 0.02 & 1\\
\hline
\end{tabular}
}
\caption{\label{tab:expected_yields} Expected event yields for signal and
backgrounds, and the observed data in all signal and control regions (marked
with an asterisk) used in the fit to extract the \ttW and \ttZ cross sections.
The quoted uncertainties on expected event yields represent systematic
uncertainties including MC statistical uncertainties.  The \ttbar, single-top,
$tZ$, $WtZ$, \ttH and $\ttbar \ttbar$ processes are denoted $t+X$. The $Z$,
$WW$, $WZ$, $ZZ$, $\ttbar WW$ and $W^{\pm} W^{\pm}$ processes are denoted
`Bosons'.}
\end{table}

\subsection{Same-sign dilepton channel}
\label{sec:SSLC}

The \SSLC\ targets the \ttW process.  Events are required to have
$\met>\SI{40}{\gev}$, $\HT>\SI{240}{\gev}$ and to contain at least two
$b$-tagged jets.  The \SSLC\ is divided into three orthogonal regions based on
the flavour combination of the lepton pair: \SSLSRA, \SSLSRB, and \SSLSRC.  In
the \SSLSRA\ region, an additional requirement on the dilepton mass removing
events with $75 \gev<m_{ee}<105 \gev$ is imposed to reduce the contamination by
$Z\to ee$ events where the charge of one electron is misidentified. A similar
requirement is not imposed on the \SSLSRB\ or \SSLSRC\ regions, since the
probability for the muon charge to be misidentified is found to be negligible,
and $Z$+jets is not a dominant background in the $e\mu$ region.  

Signal events from the \ttW process are produced when the associated $W$ boson
decays leptonically and the \ttbar system decays in the $\ell+$jets channel.

A smaller contribution from \ttZ comes from a leptonic decay of the $Z$ boson
where one lepton is not reconstructed, together with a leptonic decay of one of
the two $W$ bosons coming from the top quark decays.

The main backgrounds vary depending on the lepton flavour: events containing a
lepton with mis-identified charge are dominant in the \SSLSRA\ region and
prevalent in the \SSLSRB\ region, whereas events with a fake lepton contribute
significantly in all regions, but are dominant in the \SSLSRC\ region.
Backgrounds from the production of prompt leptons with correctly identified
charge come primarily from $WZ$ production, but these are small compared to the
instrumental backgrounds.

Processes featuring an opposite-sign lepton pair, like \ttbar and $Z$ boson
production, can enter this channel through the misidentification of the
electron charge.  Charge misidentification rates, parameterised in \pT and
$|\eta|$ of the electrons, are measured in a control region containing events
with two electrons with $75<m_{ee}<\SI{105}{\gev}$, which is divided into
same-sign and opposite-sign subregions with non-$Z\to ee$ backgrounds
subtracted. A likelihood function is constructed relating the number of
observed events in the two subregions with the probability for an electron
falling in a given $(\pT, |\eta|)$ bin to be reconstructed with the wrong
charge, and maximised to obtain the charge misidentification rates.

A template is then constructed using opposite-sign data with event selection
identical to that used in the signal region except for the requirement on the
charge of the leptons. A weight given by
\begin{equation} w = \frac{\varepsilon_{1}+ \varepsilon_{2}-
  2\varepsilon_{1}\varepsilon_{2}}
    {1 - (\varepsilon_{1}+ \varepsilon_{2}- 2\varepsilon_{1}\varepsilon_{2})}
\end{equation}
is applied to each event and used to construct the template, where the charge
misidentification rates for the two leptons are $\varepsilon_{1,2}$. These are
set to zero in the case of muons.

\begin{figure}[htbp]
\centering
\subfigure[]{\includegraphics[width=0.325\textwidth]{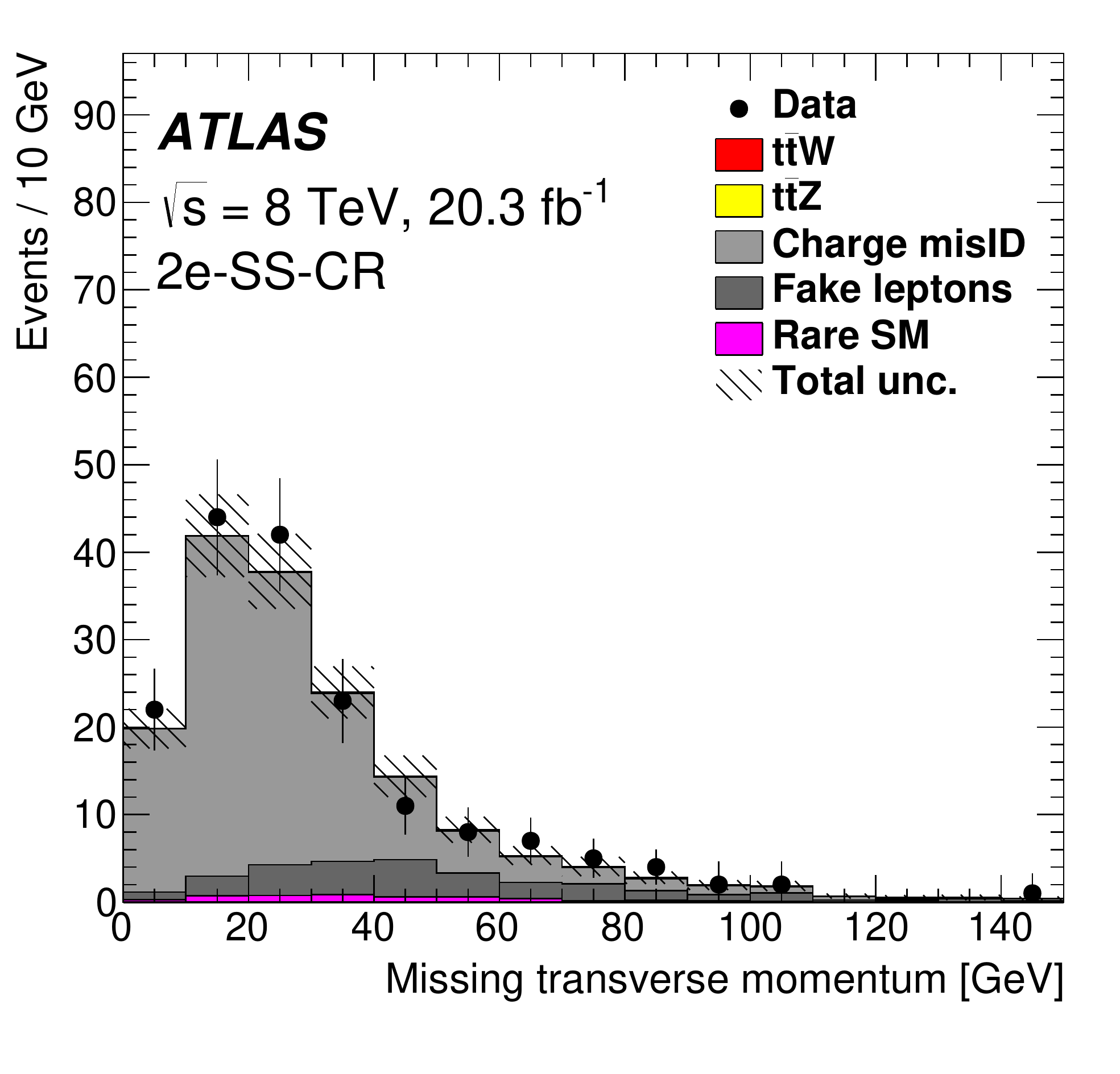}}
\subfigure[]{\includegraphics[width=0.325\textwidth]{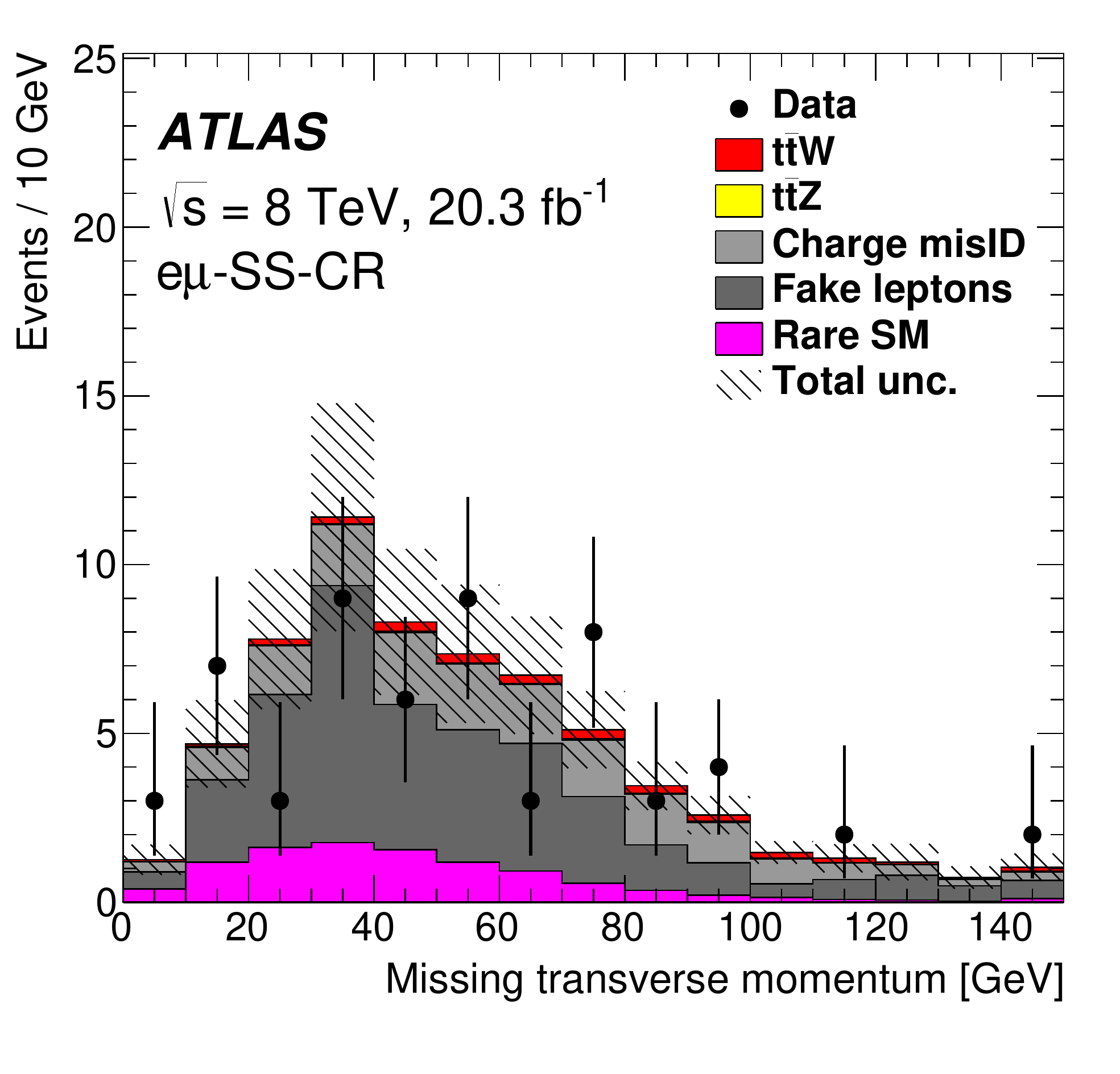}}
\subfigure[]{\includegraphics[width=0.325\textwidth]{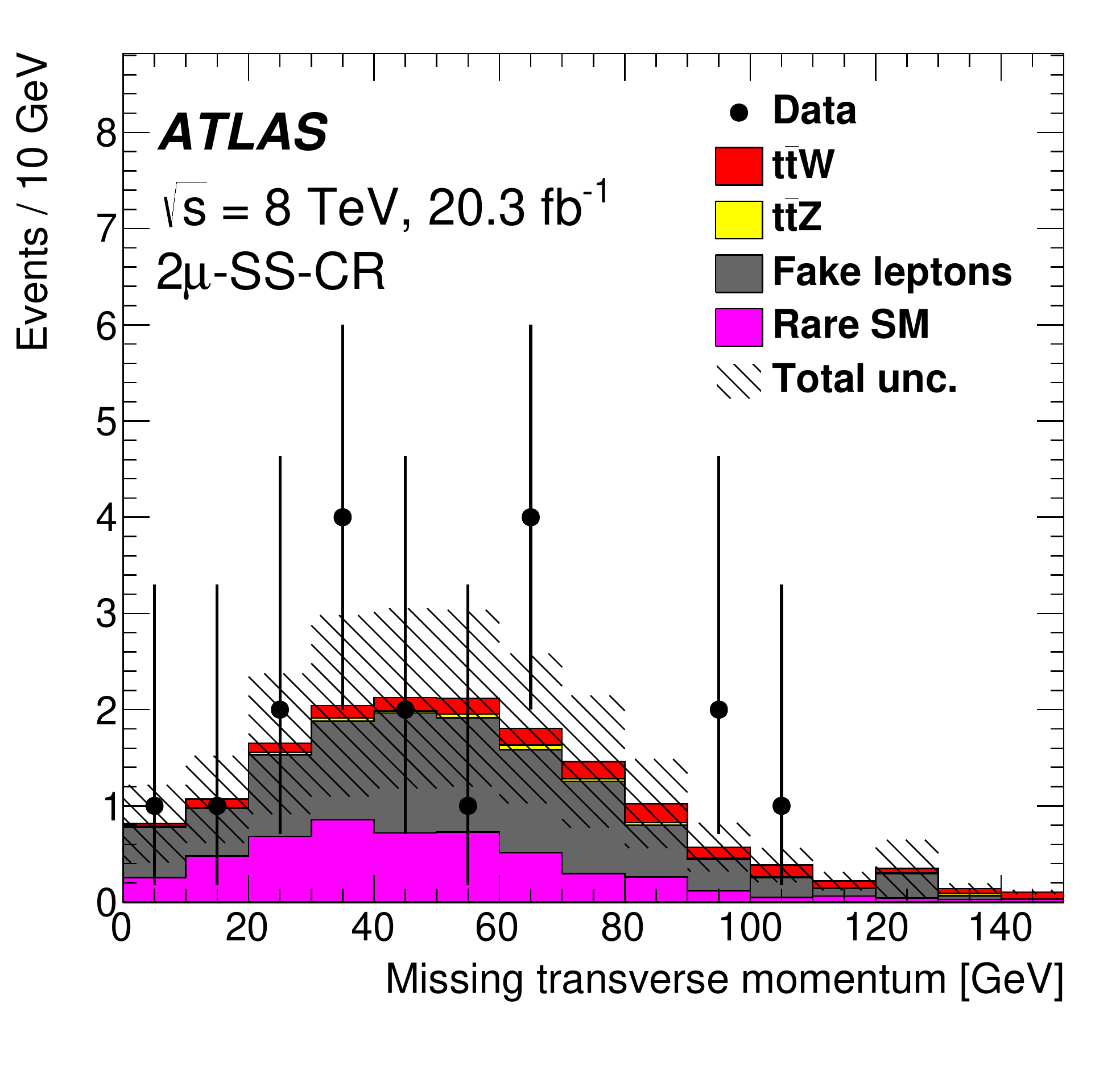}}
\caption{
Distributions of \met for events in a \SSL\ control region with
$\HT<\SI{240}{\gev}$ and at least one $b$-tagged jet for the different lepton
flavour combinations, (a) $ee$,  (b) $e\mu$ and (c) $\mu\mu$.  ``Rare SM''
contains small background contributions mainly consisting of the  $WW$ and $WZ$
processes.  The predictions are shown before fitting to data in the control
region. The instrumental backgrounds, including fake leptons and leptons with
misidentified charge are predicted using data-driven methods.   The hatched
area corresponds to the total uncertainty on the predicted yields. The last bin
in each histogram includes the overflow.}
\label{fig:SS-plots-bkg}
\end{figure}

To estimate the background from fake leptons in the \SSLC\, a set of scale
factors are measured.  The scale factors are defined as
$f=N_\mathrm{T}/N_\mathrm{L}$, the ratio of the number of observed tight
leptons, i.e.  leptons satisfying all selection criteria, to the number of
loose leptons.  Loose leptons differ from tight leptons in that they are
required to fail isolation requirements; loose muons additionally have relaxed
selection criteria, requiring $|z_0|<\SI{2}{mm}$ with no requirement on $d_0$.
The scale factors are measured in a control region (2$\ell$-SS-CR) containing
two same-sign leptons (vetoing events with a third lepton, as is done in the
signal region), at least one $b$-tagged jet, and $\HT< \SI{240}{\gev}$.  The
missing transverse momentum distributions in this control region are shown in
\Fig{SS-plots-bkg}.
A template for fake lepton backgrounds is constructed in the control region
from the loose lepton sample, using the expression
\begin{equation} \label{eqn:ff_yield}
  N_\mathrm{fake} = \left[\sum_{N_\mathrm{LT}} f_1 + \sum_{N_\mathrm{TL}} 
  f_2 - \sum_{N_\mathrm{LL}} f_1 f_2\right]_\mathrm{data} - 
  \left[ \sum_{N_\mathrm{LT}} f_1 + \sum_{N_\mathrm{TL}} f_2 - 
  \sum_{N_\mathrm{LL}} f_1 f_2 \right]_\mathrm{MC,prompt},
\end{equation}
where $N_\mathrm{TL}$ is the number of events in which the first lepton is a
tight lepton and the second is loose, and $N_\mathrm{LT}$, $N_\mathrm{LL}$
are defined in a similar fashion.  The $f_{1,2}$ are the scale factors for the
first and second leptons.  A subtracted term, shown in the brackets labelled
``MC, prompt'' is included to remove contamination from prompt lepton
production in the loose lepton sample.  This subtraction accounts for about
2--3\% of the total estimate.  The background template for fake leptons is
fitted to data in the control region in bins of lepton \pt to obtain the scale
factors $f_i$, which are measured separately for electrons and muons and also
binned in lepton \pt.  With the measured scale factors, the background
templates are produced in the signal region according to
Eq.~(\ref{eqn:ff_yield}).

Potential overlap between the estimates of charge misidentification and fake
leptons is taken into account with an additional subtraction step.  A
background template for fake leptons is produced for opposite-sign events
(using a selection and binning otherwise the same as the signal region), and
the charge misidentification rates are applied to this template to obtain a
representation of the overlap of these two backgrounds.  This new template is
then subtracted from the total estimate of the fake lepton background. This
subtraction represents about 2\%--5\% of the total yield.

\begin{figure}[htbp]
\centering
\subfigure[]{\includegraphics[width=0.275\textwidth]{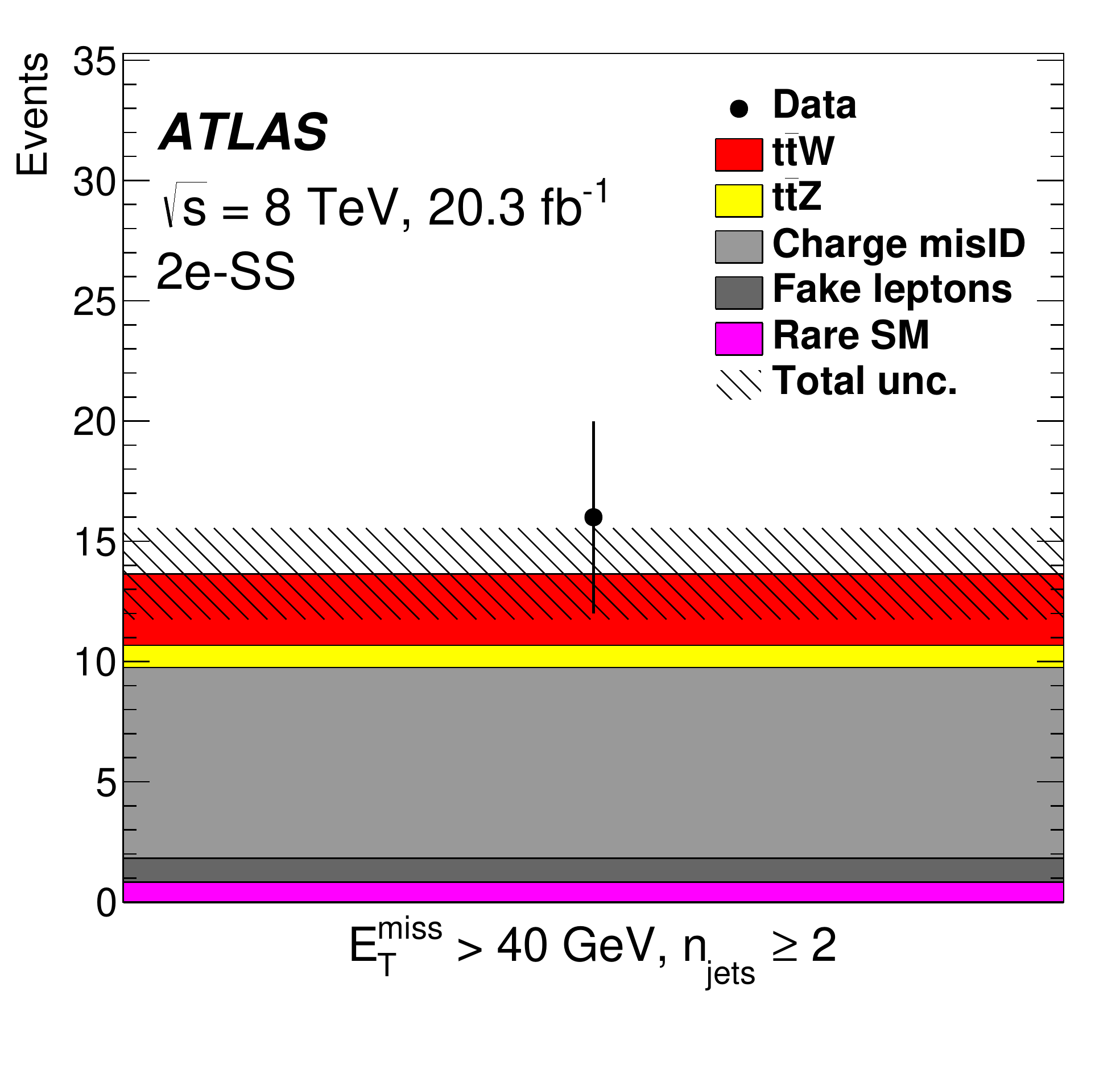}}
\subfigure[]{\includegraphics[width=0.355\textwidth]{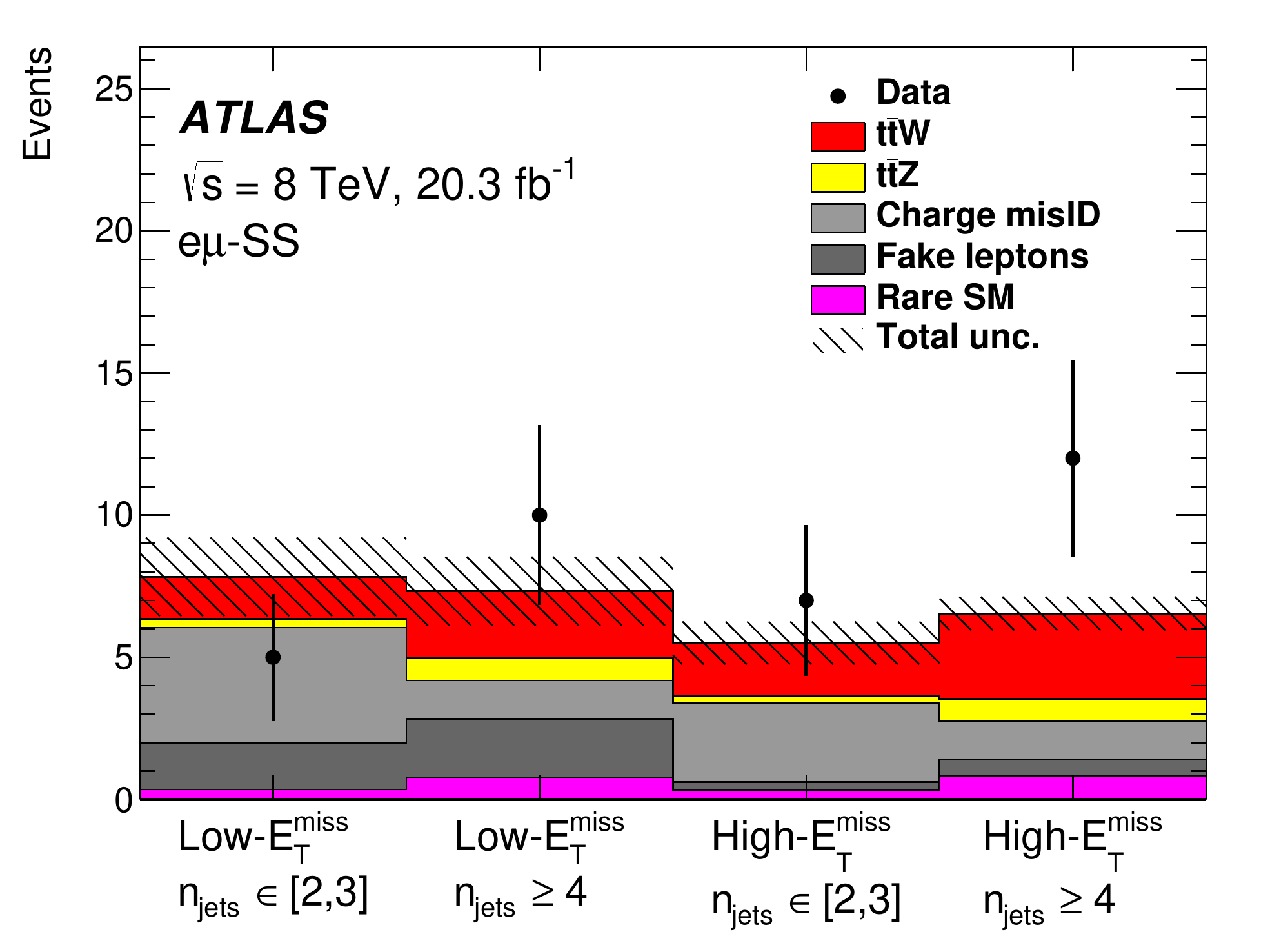}}
\subfigure[]{\includegraphics[width=0.355\textwidth]{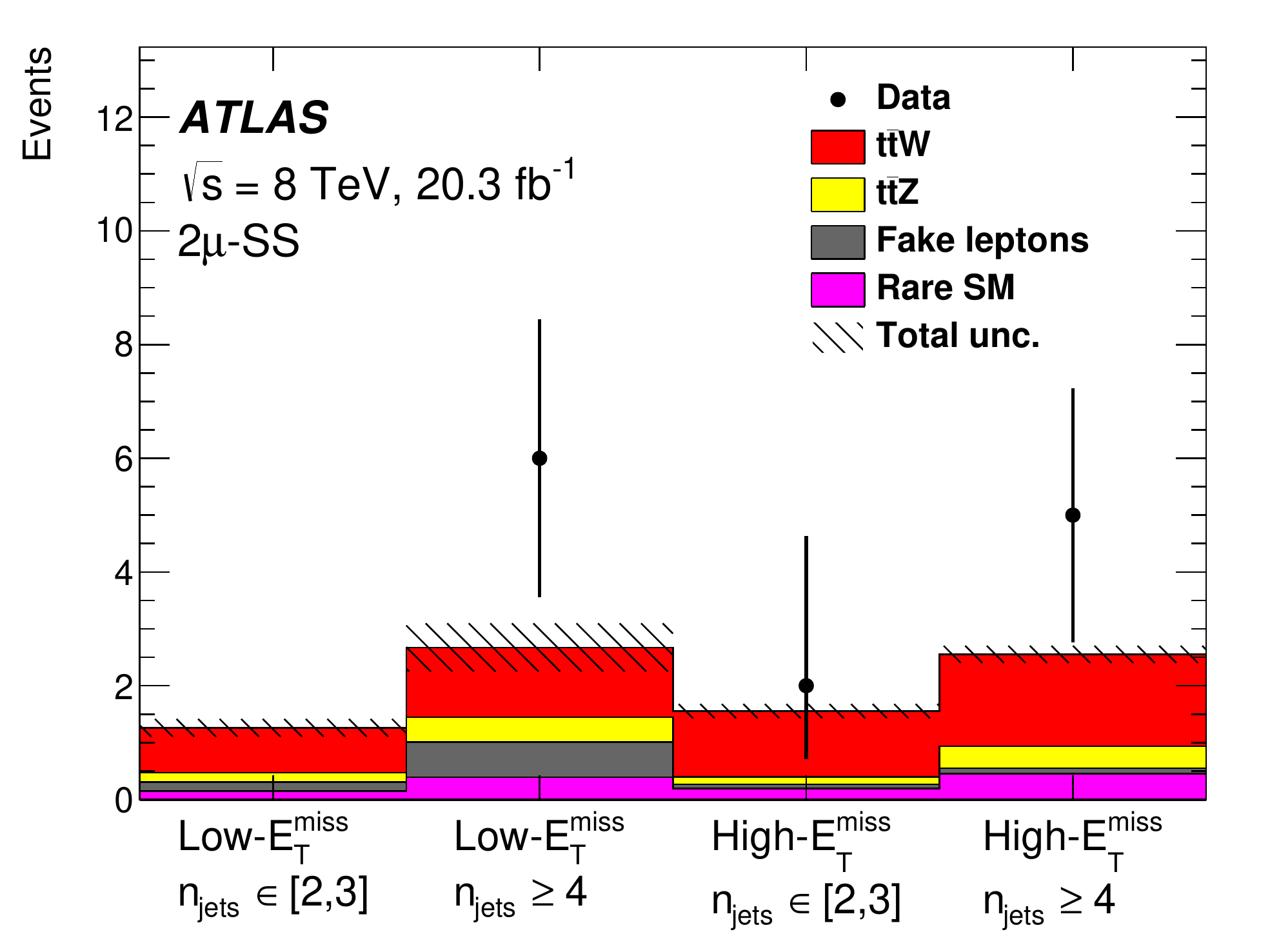}}
\caption{ 
Event yields in the \SSL\ signal regions according to the binning used in the
final likelihood fit in the (a) \SSLSRA\ (b) \SSLSRB\ and (c) \SSLSRC\ regions.
The distributions are shown before the fit. The bins labelled ``Low-\met''
correspond to $\met \in (40,80) \gev$, and those labelled ``High-\met''
correspond to $\met \geq \SI{80}{\gev}$.  ``Rare SM'' contains small background
contributions mainly consisting of the \ttH and $WZ$ processes.
Instrumental backgrounds, including fake leptons and leptons with misidentified
charge are predicted using data-driven methods.  The hatched area corresponds
to the total uncertainty on the predicted yields.}
\label{fig:SS-plots}
\end{figure}

To improve the separation of the \ttW signal from backgrounds, events in the
\SSLSRB\ and \SSLSRC\ regions are further divided into four bins based on jet
multiplicity and missing transverse momentum.  Events are classified as
low-\njets\ ($\njets=2$ or 3) or high-\njets\ ($\njets \geq 4$) and low-\met
($40<\met < \SI{80}{\gev}$) or high-\met ($\met \ge 80$~\gev).  No further
event classification is used in the \SSLSRA\ region. The expected and observed
contributions in each of the three dilepton flavour regions are summarised in
\Tab{expected_yields} and plotted in \Fig{SS-plots}.

\subsection{Trilepton channel}
\label{sec:TLC}

In the \TLC\ two preselections are considered, referred to as \threez\ and
\threezveto. The \threez\ region targets the \ttZ process, while \threezveto\
aims at measuring the \ttW process. In the region \threez, at least one pair of
leptons is required to have the opposite sign and same flavour (OSSF) and have
an invariant mass within \SI{10}{\gev} of the $Z$ boson mass. Region
\threezveto\ contains the remaining \TL\ events with a requirement that the
leptons must not all have the same sign.

The \TL\ channel signal regions are determined as follows.  First, the
preselected samples are split into categories according to the jet multiplicity
and the number of $b$-tagged jets. The categories with similar predicted
signal-to-background ratio ($S/B$) and systematic uncertainties are grouped
together.  The final selection in each group is optimised for maximal expected
significance, including both the statistical and systematic uncertainties,
using requirements on $\met$ and lepton \pT.  It is found that optimal
significance is obtained without a requirement on $\met$. 

Four signal regions are defined as a result of the grouping and optimisation:
\TLSRA, \TLSRB, \TLSRC\ and \TLSRD.  In the \TLSRA\ region, at least four jets
are required, exactly one of which is $b$-tagged.  In the \TLSRB\ region,
exactly three jets with at least two $b$-tagged jets are required. In the
\TLSRC\ region at least four jets are required, of which at least two jets are
$b$-tagged. In the \TLSRD\ region at least two and at most three jets are
required, of which at least two jets are $b$-tagged. For events in which the
third leading lepton is an electron, the minimum \pT requirement on the third lepton is
raised to \SI{20}{\gev} in the \TLSRA, \TLSRB\ and \TLSRC\ regions, and to
\SI{25}{\gev} in the \TLSRD\ region. 

\label{sec:wz}

The \threez\ preselection is dominated by $WZ$ events, with a significant
contribution from events with fake leptons.  To constrain the $WZ$ background,
a control region called \TLCR\ is defined and included in the fit. In this
region, the presence of exactly three jets, with exactly zero $b$-tags, is
required in addition to the requirements of the \threez\ preselection. The
normalisation correction for the $WZ$ background with respect to the Standard
Model expectations is obtained from the fit and found to be $0.98 \pm 0.20$.
The quoted uncertainty includes both the statistical and systematic components.
The modelling of $WZ$ production in association with heavy-flavour jets is
further validated in a control region \mbox{3$\ell$-Z-1b-CR}, defined by
requiring the presence of one to three jets, exactly one of which is
$b$-tagged.

\begin{figure}[htbp]
\centering
\subfigure[]{\includegraphics[width=0.495\textwidth]{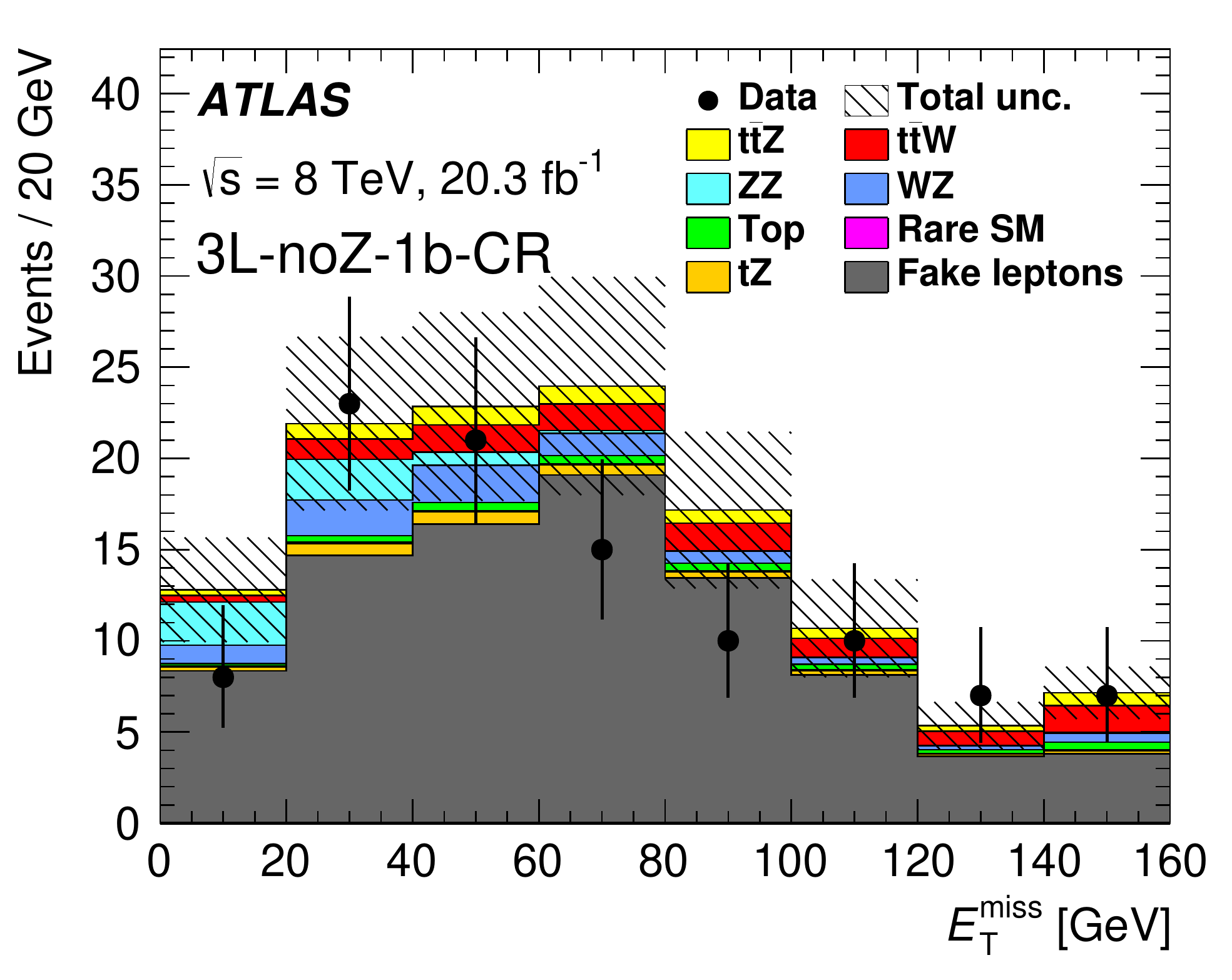}}
\subfigure[]{\includegraphics[width=0.495\textwidth]{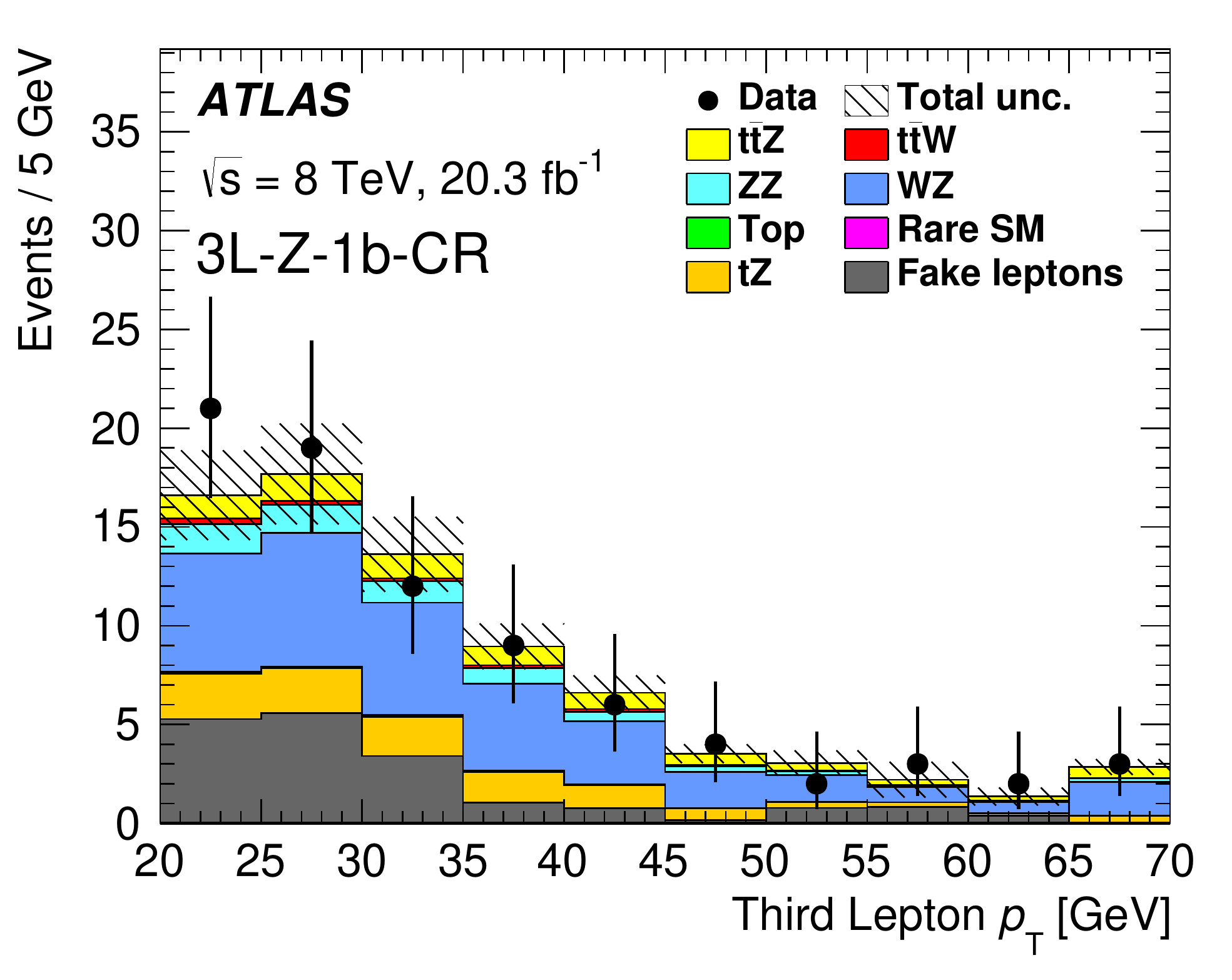}}
\caption{Distributions of (a) \met in the \mbox{3$\ell$-noZ-1b} region and (b)
third-lepton \pT in the \mbox{3$\ell$-Z-1b} region. ``Rare SM'' contains small
background contributions consisting of the  $WWW$, $WWZ$, $H \rightarrow ZZ$
and $t\bar{t}WW$ processes.  The hatched area corresponds to the total
uncertainty on the predicted yields. The distributions are shown before the
fit. The last bin in each histogram includes the overflow.}
\label{fig:CRZ}
\end{figure}

The fake lepton background is estimated by using the so-called matrix
method~\cite{Aad:2010ey}, which makes use of an orthogonal control region in
which lepton isolation and electron identification criteria are relaxed. The
efficiencies for real and fake leptons used in the matrix method are measured
in events containing two leptons and one $b$-tagged jet.  To validate the
estimate of the background containing fake leptons, a control region
\mbox{3$\ell$-noZ-1b-CR} is defined by requiring exactly one jet to be
$b$-tagged in addition to the requirements of the \threezveto\ region.
\Fig{CRZ} shows distributions of \met and third-lepton \pT in the
\mbox{3$\ell$-noZ-1b-CR} and \mbox{3$\ell$-Z-1b-CR} regions, respectively.  The
level of agreement between data and expectation is good.
 
\begin{figure}[htbp]
\centering
\subfigure[]{\includegraphics[width=0.495\textwidth]{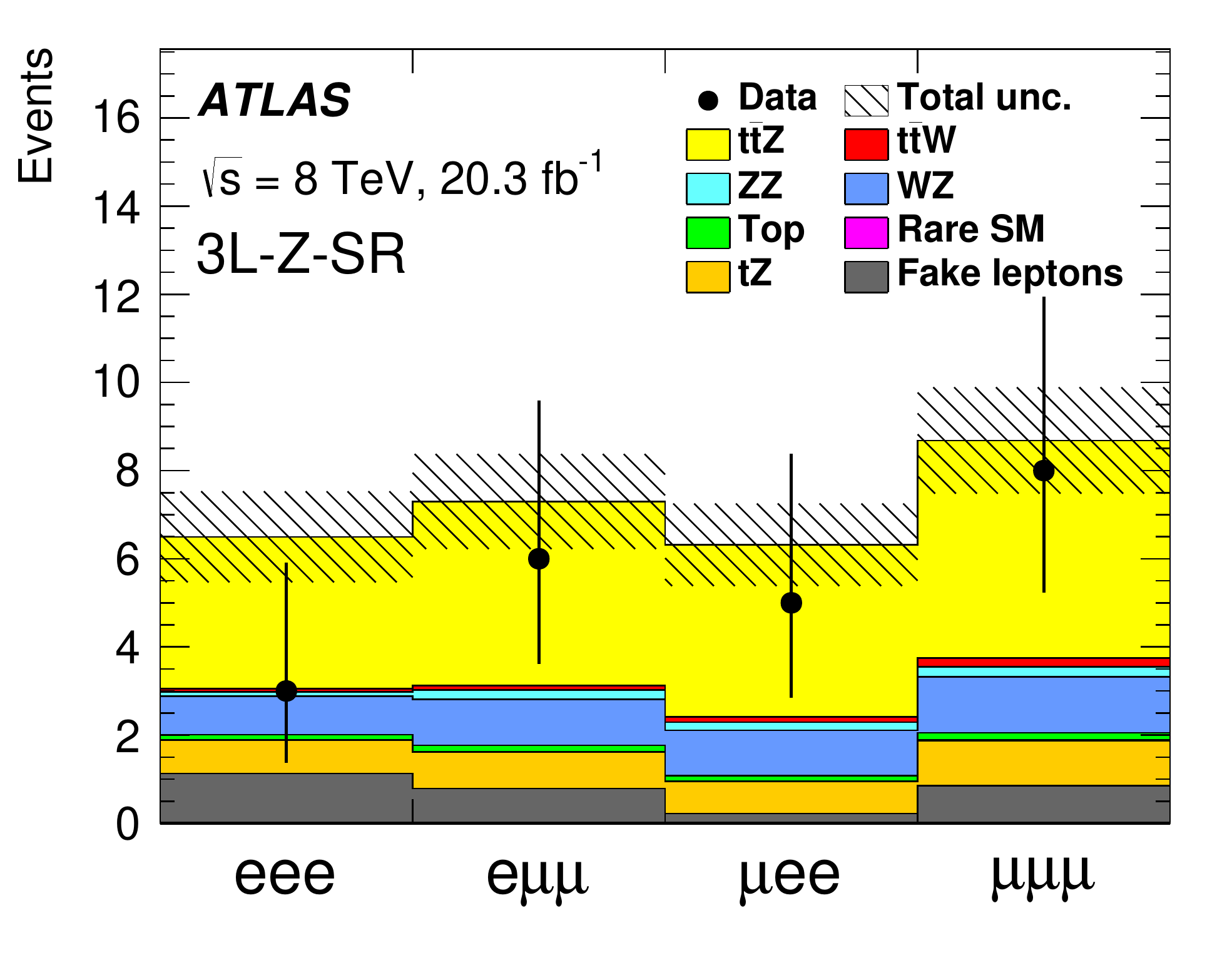}}
\subfigure[]{\includegraphics[width=0.495\textwidth]{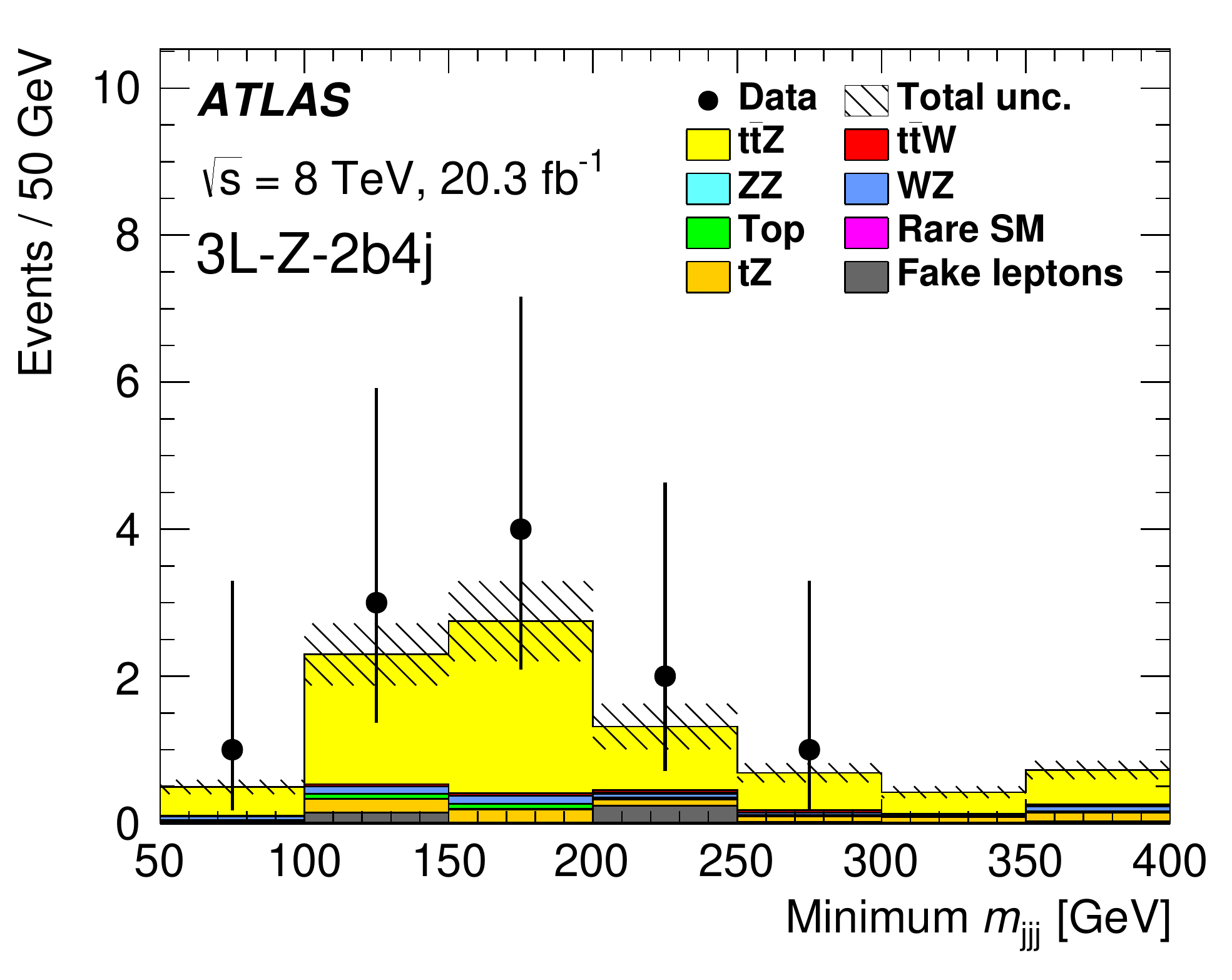}}
\caption{(a) Event yields in the \TLC\ summarising the signal regions with
different lepton flavour combinations and (b) the minimum three-jet invariant
mass for events in the \TLSRC\ signal region.  ``Rare SM'' contains small
background contributions consisting of the $WWW$, $WWZ$, $H \rightarrow ZZ$ and
$t\bar{t}WW$ processes.  The distributions are shown before the fit. The
hatched area corresponds to the total uncertainty on the predicted yields. The
last bin in (b) includes the overflow. }
\label{fig:SR3l} 
\end{figure}

The signal and control regions of the \TLC\ used in the fit are summarised in
Table~\ref{tab:TLregions}.  The expected and observed yields in the signal and
control regions are shown in \Tab{expected_yields}. Event yields summarising
the signal regions with different lepton flavour combinations (3$\ell$-Z-SR)
and the distribution of the minimum invariant mass of jet triplets (minimum
$m_{jjj}$) for events in the \TLSRC\ region are shown in \Fig{SR3l}.
Considering the four leading jets, the momentum vector sum of the minimum
invariant jet triplet mass is found to give a powerful estimate of the
hadronically decaying top direction. Good agreement between data and
expectation is observed. 

\begin{table}
\centering \renewcommand{\arraystretch}{1.2}
\begin{tabular}{|ccc|}
\hline
Region & Targeting & Sample fraction [\%] \\
\hline
\hline
3$\ell$-Z-1b4j & \multirow{3}{*}{\ttZ}  & 47\\
3$\ell$-Z-2b3j & & 54 \\
3$\ell$-Z-2b4j & & 76 \\ \hline
3$\ell$-noZ-2b & $\ttW$ &  48 \\
3$\ell$-Z-0b3j  & $WZ$ & 68\\
\hline
\end{tabular}
\caption{\label{tab:TLregions} Signal and control regions of the \TLC\ used in
the fit, together with the processes targeted and the expected fraction of the
sample represented by the targeted process.}
\end{table}

\subsection{Tetralepton channel}
\label{sec:FLC}

The \FLC\ targets the \ttZ process for the case where both $W$ bosons resulting
from top quark decays and the $Z$ boson decay leptonically, and uses an event
counting approach in five signal regions.  Events with two pairs of
opposite-sign leptons are selected, among which at least one pair is same
flavour. The OSSF lepton pair with reconstructed invariant mass closest to
$m_Z$ is attributed to the \Zboson boson decay and denoted in the following as
$Z_1$.  The two remaining leptons are used to define $Z_2$.  The signal regions
are defined according to the relative flavour of the two remaining leptons,
different or same flavour, and the number of $b$-tagged jets: zero, one, or at
least two ($0b$, $1b$, $2b$).  The signal regions are thus \FLSRA, \FLSRB,
\FLSRC, \FLSRD\ and \FLSRE. The $ZZ$ background mostly affects the SF regions
and therefore events with a $Z_2$ SF lepton pair and no $b$-tagged jets are
discarded. 

Further requirements are applied in each signal region such that the expected
statistical uncertainty on the measured \ttZ signal cross section is minimised.
Events in the \FLSRD\ region are rejected if they are compatible with a $ZZ$
event, i.e. if \mbox{$\met<80\,(40) \gev$} for $m_{Z_2}$ inside (outside) a
\SI{10}{\gev} region centred at the $Z$ boson mass. This requirement on \met is
relaxed by \SI{40}{\gev} for the \FLSRE\ region.  The impact of events with
fake leptons decreases with the number of reconstructed \mbox{$b$-tagged} jets.
To suppress these backgrounds, additional requirements on the scalar sum of the
transverse momenta of the third and fourth leptons ($\pttf$) are imposed in the
lower $b$-tag multiplicity regions. In the \FLSRD, \FLSRB\ and \FLSRA\ regions
events are required to satisfy $\pttf > \SI{25}{\gev}$, $\pttf > \SI{35}{\gev}$
and $\pttf > \SI{45}{\gev}$, respectively.  In the \FLSRA\ region the
requirement on the fourth lepton is raised to $\pT > \SI{10}{\gev}$ and at
least two jets must be reconstructed in the event.  In all regions, the
invariant mass of any two reconstructed OS leptons is required to be larger
than \SI{10}{\gev}.  The definitions of the signal regions are summarised in
\Tab{FL-cuts}.

\begin{table}[htbp]
\centering \renewcommand{\arraystretch}{1.2}
\resizebox{\columnwidth}{!}{
\begin{tabular}{|lcccr@{}cc@{}lcc|}
\hline
Region & $Z_2$ leptons & $p_{\text{T}4}$ & \pttf && $|m_{\ell\ell} - m_{Z_{2}}|
$ & \met && $N_{\text{jets}}$ & $N_{b\text{-jets}}$\\
\hline
\hline
\FLSRA & $e^{\pm}\mu^{\mp}$ & $>\SI{10}{\gev}$ & $>\SI{45}{\gev}$ &&-&-&&
$\ge2$ & 0 \\
\FLSRB & $e^{\pm}\mu^{\mp}$ & $>\SI{7}{\gev}$ & $>\SI{35}{\gev}$ &&-&-&&-& 1\\
\FLSRC & $e^{\pm}\mu^{\mp}$ & $>\SI{7}{\gev}$ &-&&-&-&&-& $\ge2$\\
\FLSRD & $e^{\pm}e^{\mp},\mu^{\pm}\mu^{\mp}$ & $>\SI{7}{\gev}$ &
$>\SI{25}{\gev}$ &\TabLeftBracket&
\begin{tabular}{c}$>\SI{10}{\gev}$\\$<\SI{10}{\gev}$\end{tabular} &
\begin{tabular}{c}$>\SI{40}{\gev}$\\$>\SI{80}{\gev}$\end{tabular}
&\TabRightBracket&-& 1\\
\FLSRE & $e^{\pm}e^{\mp},\mu^{\pm}\mu^{\mp}$ & $>\SI{7}{\gev}$
&-&\TabLeftBracket&
\begin{tabular}{c}$>\SI{10}{\gev}$\\$<\SI{10}{\gev}$\end{tabular} &
\begin{tabular}{c}-\\$>\SI{40}{\gev}$\end{tabular} &\TabRightBracket&-& $\ge2$
\\
\hline
\end{tabular}
}
\caption{\label{tab:FL-cuts} Definitions of the five signal regions in the
\FLC.}
\end{table}

\label{sec:zz}

\begin{figure}[htbp]
\centering
\subfigure[]{\includegraphics[width=0.495\textwidth]{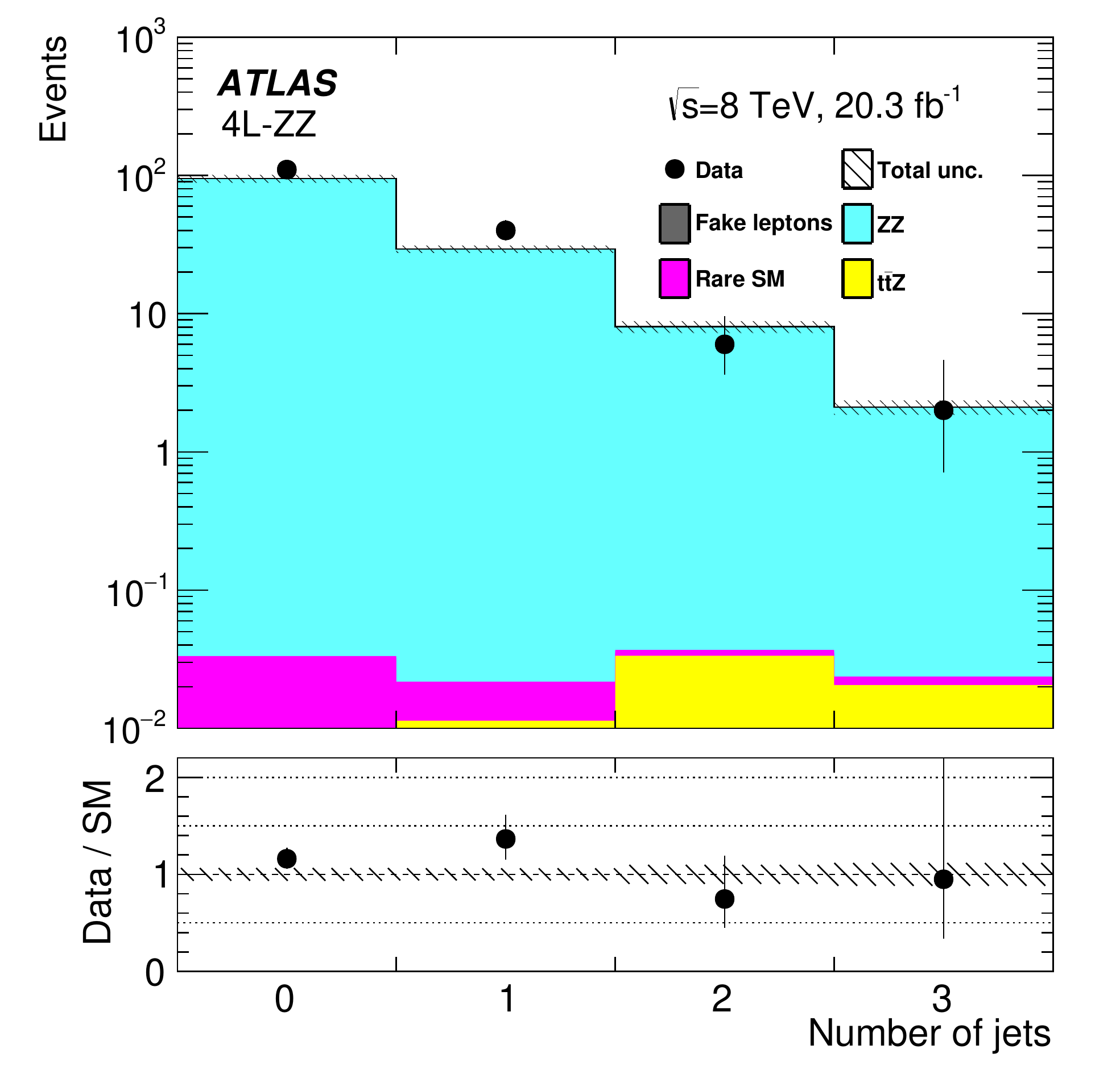}}
\subfigure[]{\includegraphics[width=0.495\textwidth]{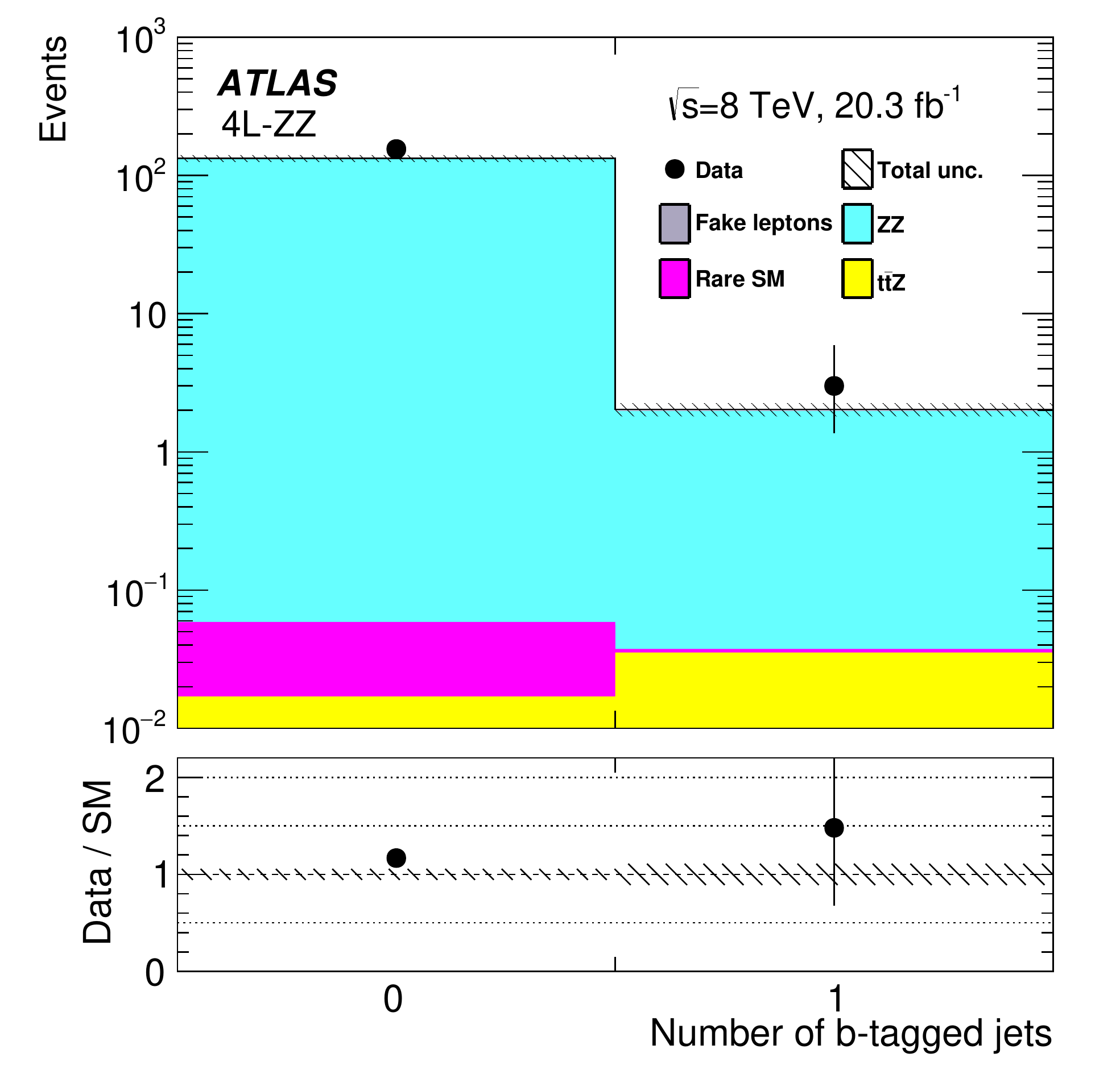}}
\caption{ Distributions of (a) the number of jets and (b) the number of
$b$-tagged jets in the $ZZ$ control region of the tetralepton channel \FLCR.
The hatched area corresponds to the total uncertainty on the predicted yields.
The distributions are shown before the fit.  ``Rare SM'' contains small
background contributions mainly consisting of the $WtZ$ and \ttH processes.
The ``Data/SM'' plots show the ratio of the data events to the total Standard
Model expectation. } 
\label{fig:fourleptonZZ} 
\end{figure}

The $ZZ$ background is large in the \FLC, and therefore a control region,
\FLCR, is defined to constrain the $ZZ$ normalisation in the SF region, and is
included in the fit. Both lepton pairs are required to satisfy $|m_{Z_{1,2}} -
m_Z| < 10 \gev$, and events are retained if $\met<50 \gev$.  The fitted
normalisation correction with respect to the Standard Model expectation is
$1.16 \pm 0.12$.  The quoted uncertainties include both the statistical and
systematic components.  The number of jets and $b$-tagged jets in the \FLCR\
region are shown in \Fig{fourleptonZZ}.  Data distributions agree with
expectations from simulation.

The \FLC\ backgrounds with at least one fake lepton are estimated using
simulation, where the prediction is corrected with a constant factor to improve
agreement in control regions with enhanced fraction of single fake leptons.  By
probing the fake muon and electron background in a $Z+$fake-lepton-candidate
control region and in a $\ttbar+$fake-lepton-candidate region, two equations
per lepton flavour $f$ can be constructed and the correction factors
$c^{\text{CR}}_f$ for these two processes can be determined: $c^{\ttbar}_e =
1.23 \pm 0.13$, $c^{\ttbar}_\mu = 1.25 \pm 0.09$, $c^Z_e = 1.35 \pm 0.05$, and
$c^Z_\mu = 1.61 \pm 0.05$. The quoted uncertainties include systematic effects.
The control regions are required to contain three leptons and are either $Z$-
or \ttbar-like. In the first case an OSSF lepton pair is required together with
$\met<\SI{30}{\gev}$. In addition, the transverse mass of the non-$Z_1$ lepton
$\ell$ is required to satisfy $m_\text{T} <\SI{35}{\gev}$, where $m_\text{T}$
is defined as
\begin{equation}
m_\text{T} = \sqrt{2p^{\ell}_\mathrm{T}\met - 2\mathbf{p}^{\ell}_\mathrm{T} \cdot 
\mathbf{p}^\mathrm{miss}_\mathrm{T}}.
\end{equation}
The non-$Z_1$ lepton is then used as the fake lepton candidate. In the second
case an OS lepton pair is required together with at least one jet with
$\pT>\SI{30}{\gev}$, and events with an OSSF lepton pair are rejected. The
lowest-\pT same-sign lepton is then used as the candidate.  The background from
events with two fake leptons is evaluated from simulation with relaxed
requirements and extrapolated in several steps into the signal region. The
total background yield and its uncertainty are dominated by the estimate
extracted from simulation of trilepton events with only one additional fake
lepton.

\begin{figure}[htbp]
\centering
\subfigure[]{\includegraphics[width=0.495\textwidth]{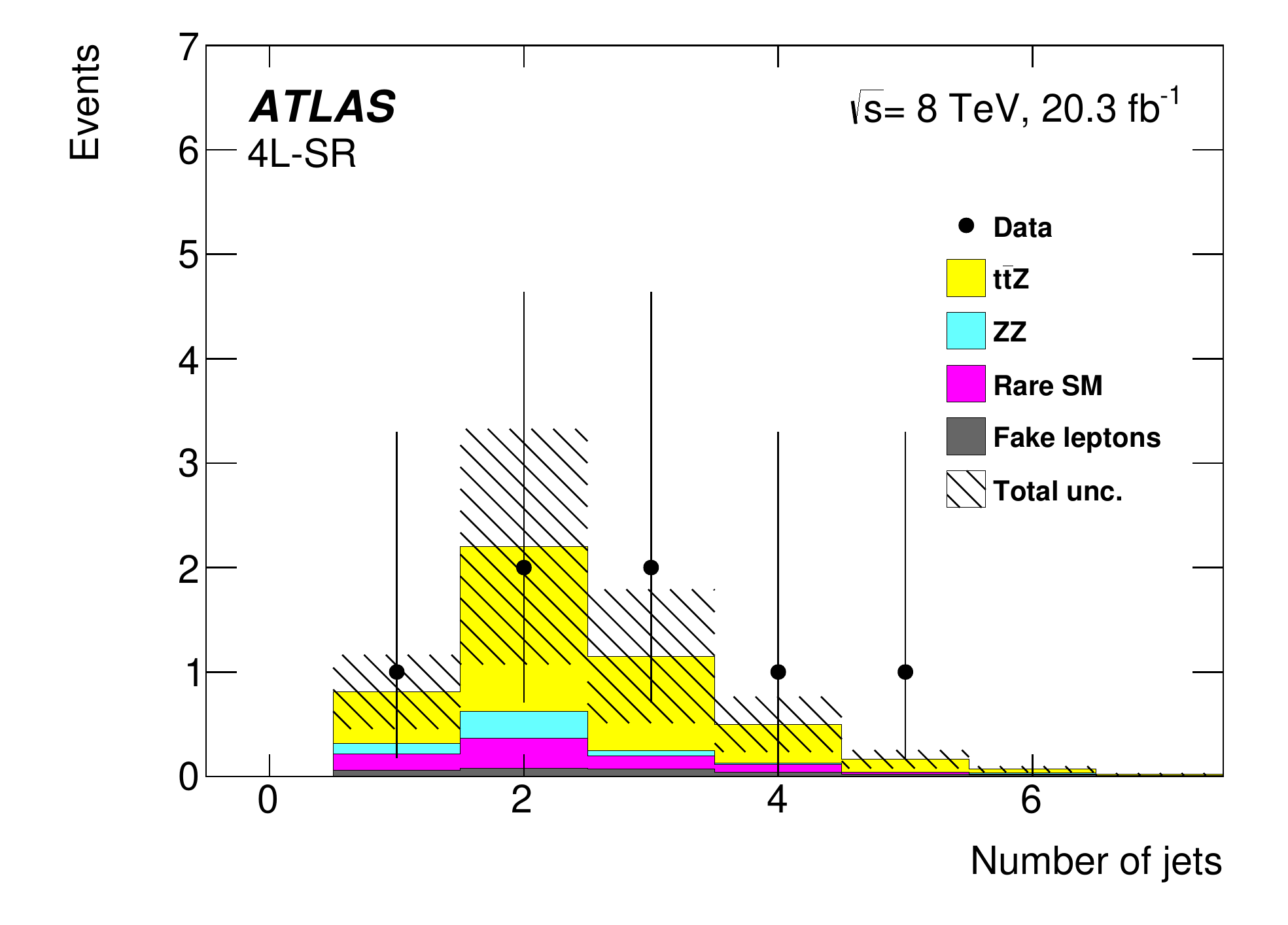}}
\subfigure[]{\includegraphics[width=0.495\textwidth]{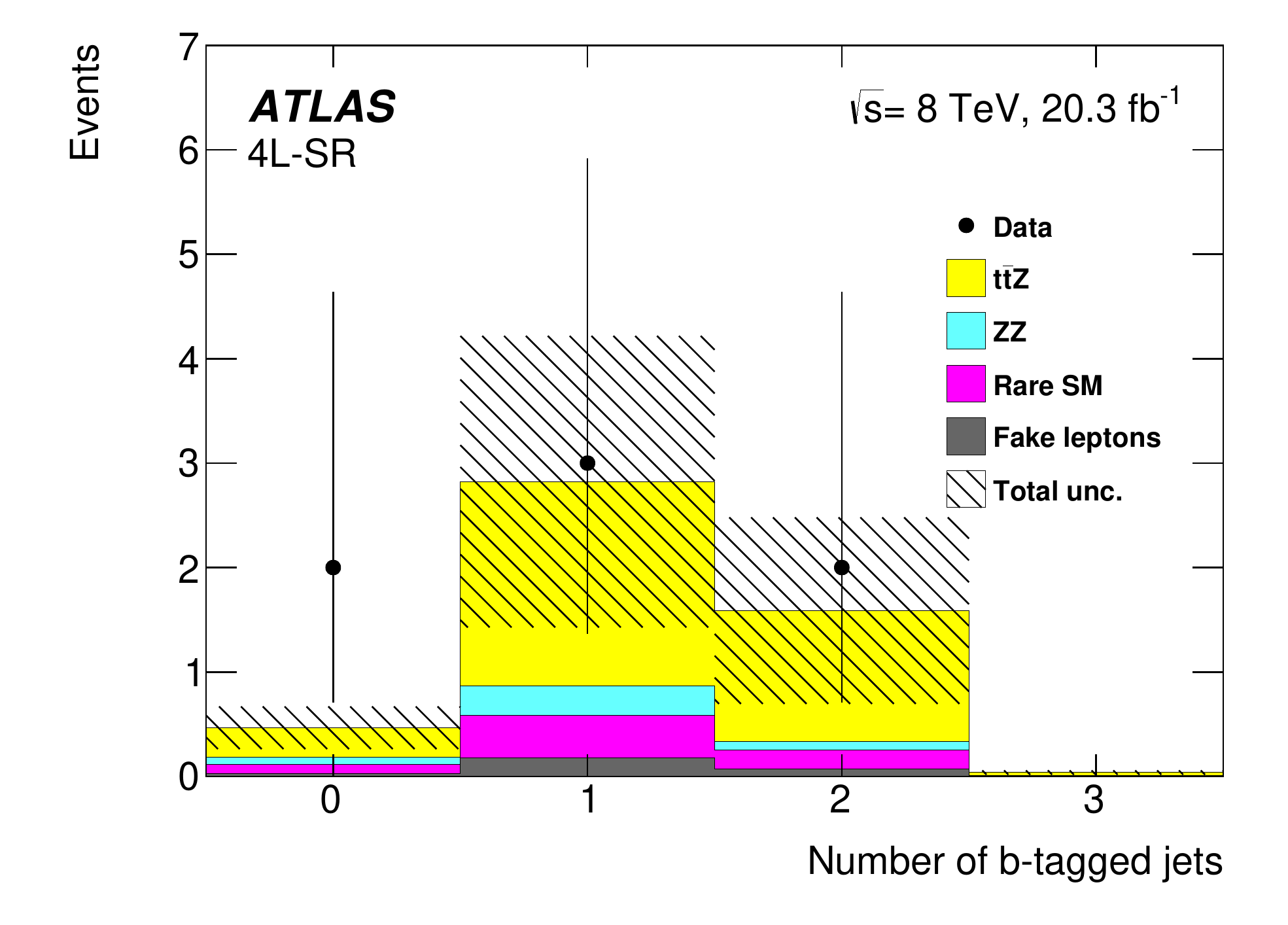}}
\subfigure[]{\includegraphics[width=0.495\textwidth]{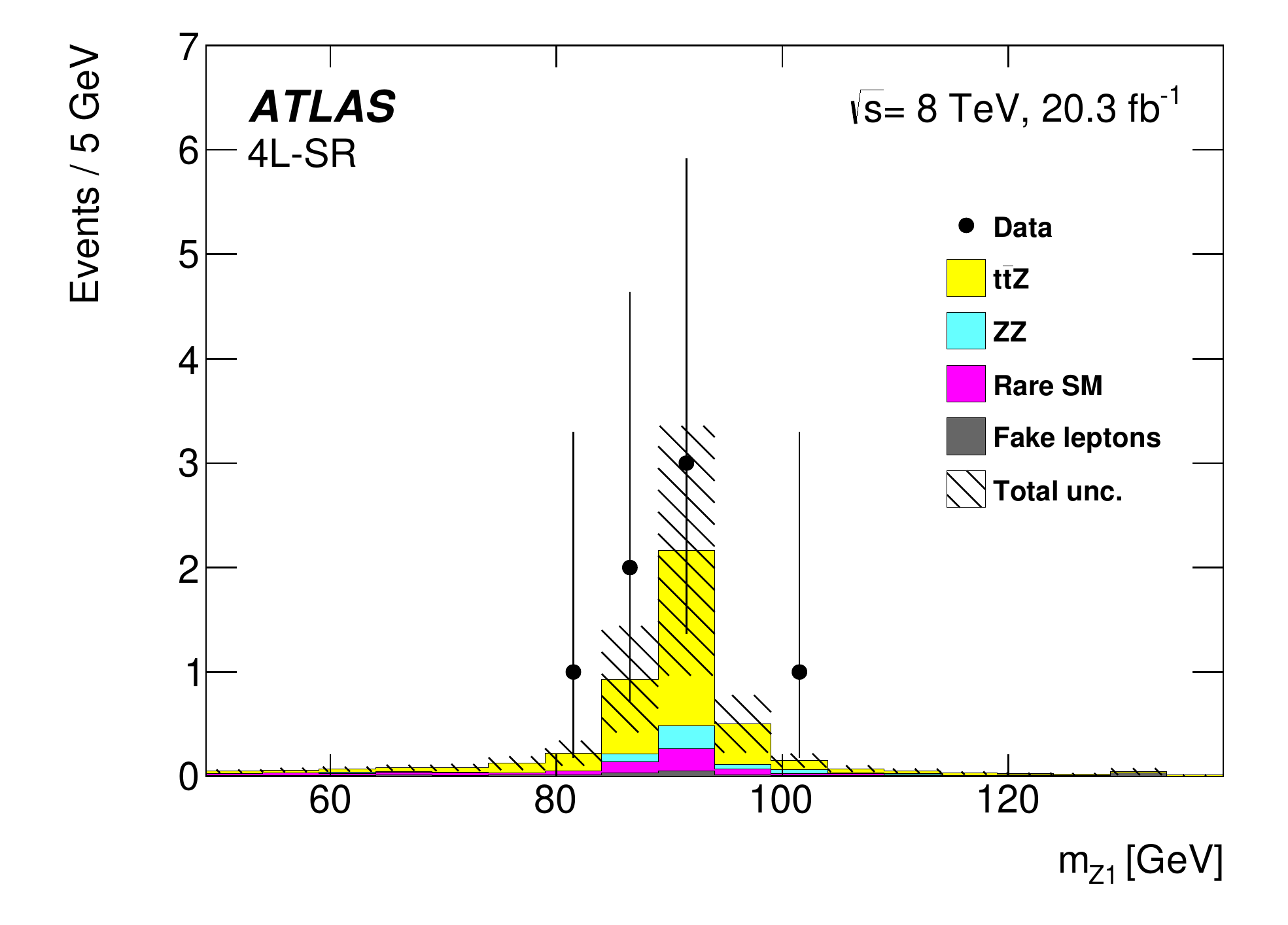}}
\subfigure[]{\includegraphics[width=0.495\textwidth]{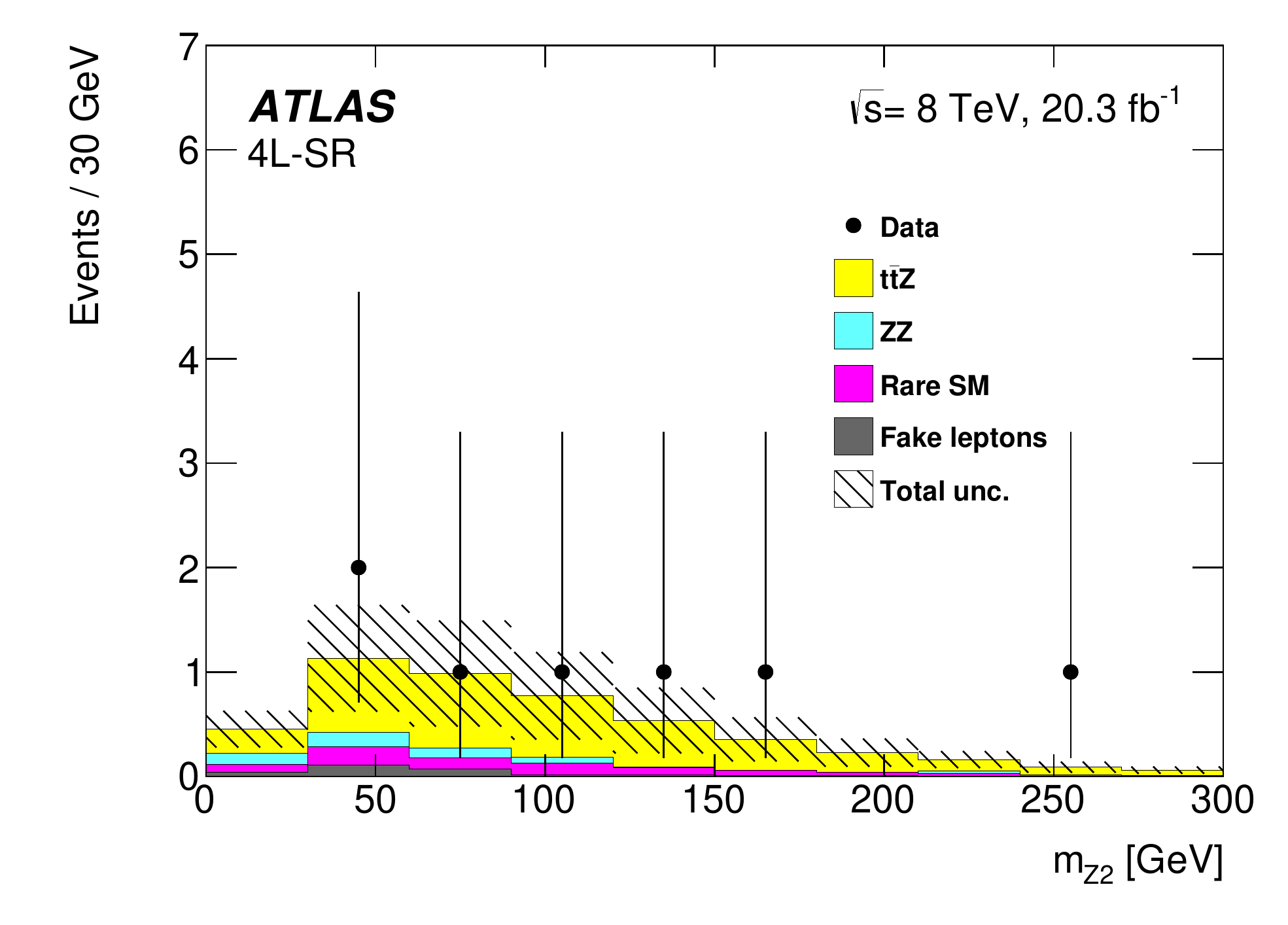}}
\caption{Distributions of (a) number of jets, (b) number of $b$-tagged jets,
invariant mass of the (c) $Z_1$ and (d) $Z_2$ dilepton pair for the \FL\ signal
region selection. The distributions are shown before the fit.  The
distributions of the seven observed events are compared to expectation. The
hatched area corresponds to the total uncertainty on the predicted yields.
``Rare SM'' contains small background contributions mainly consisting of the
$WtZ$ and \ttH processes.}
\label{fig:fourlepton}
\end{figure}

The expected sample composition of the six \FL\ regions is summarised in
\Tab{expected_yields} along with the number of events observed in data.  Seven
events are observed in the five signal regions (4L-SR).  \Fig{fourlepton} shows
good agreement between data and expectation for the distributions of the number
of jets, number of $b$-tagged jets, as well as the invariant masses of the two
pairs of leptons. 

\section{Systematic uncertainties}
\label{sec:SystematicUncertainties}				   

Several sources of systematic uncertainty are considered that can affect the
normalisation of signal and background in each channel and/or the shape of the
discriminant distributions in the \OSLC.

The luminosity estimate has an uncertainty of 2.8\%, determined using
beam-separation scans~\cite{DAPR-2011-01}.  This systematic uncertainty is
assigned to all background contributions obtained from MC simulation. 

\subsection{Uncertainties on reconstructed objects}
\label{sec:syst_objects}
 
Uncertainties associated with the lepton selection arise from the imperfect
knowledge of the trigger, reconstruction, identification and isolation
efficiencies, and lepton momentum scale and resolution. The uncertainty on the
electron identification efficiency is the largest systematic uncertainty in the
\TLC\ and among the most important ones in the \FLC.    

Uncertainties associated with the jet selection arise from the jet energy scale
(JES), JVF requirement, jet energy resolution (JER) and jet reconstruction
efficiency.  The JES and its uncertainty are derived combining information from
test-beam data, collision data and simulation~\cite{PERF-2012-01}.  JES
uncertainty components arising from the in-situ calibration and the jet flavour
composition are among the dominant uncertainties in the \OSL, \SSL\ and \TL\
channels.  The uncertainties in the JER and JVF have a significant effect at
low jet \pt. The JER uncertainty is the second largest uncertainty in the \TLC\
while the JVF uncertainty is not negligible in the \OSL, \TL\ and \FL\
channels.

The efficiency of the flavour tagging algorithm is measured for each jet
flavour using control samples in data and in simulation. From these
measurements, correction factors are defined to correct the tagging rates in
the simulation. In the case of $b$-jets, correction factors and their
uncertainties are estimated based on observed and simulated $b$-tagging rates
in \ttbar dilepton events~\cite{ATLAS-CONF-2014-004}.  In the case of $c$-jets,
they are derived based on jets with identified $D^*$
mesons~\cite{ATLAS-CONF-2014-046}.  In both cases the correction factors are
parameterised as a function of jet $\pt$. In the case of light-flavour jets,
correction factors are derived using dijet events, and are parameterised as a
function of jet $\pt$ and $\eta$~\cite{ATLAS-CONF-2014-046}.  Sources of
uncertainty affecting the $b$- and $c$-tagging efficiencies are considered as a
function of jet \pt, including bin-to-bin
correlations~\cite{ATLAS-CONF-2014-004}.  An additional uncertainty is assigned
to account for the extrapolation of the $b$-tagging efficiency measurement from
the \pT region used to determine the scale factors to regions with higher \pT.
For the light-jet tagging efficiency the dependence of the uncertainty on the
jet \pt and \eta is considered.  These systematic uncertainties are taken as
uncorrelated between $b$-jets, $c$-jets, and light-flavour jets. 

The treatment of the uncertainties on reconstructed objects is common to all
four channels, and thus these are considered as correlated among different
regions.

\subsection{Uncertainties on signal modelling} 
\label{sec:ttV_PS}

To assess the factorisation and renormalisation scale uncertainties on \ttV
modelling, the scales are varied up and down by a factor of two in a correlated
manner.  Radiation uncertainties are assessed by simultaneously varying the
scale of the momentum transfer $Q$ in the running strong coupling $\alpha_\mathrm{S}
 (Q^2)$ in the matrix-element calculation and in the \textsc{Pythia} parton
shower, up or down by a factor of two~\cite{Cooper:2011gk}.

In addition, the jet \pT matching threshold and the amount of radiation in the
parton shower are independently varied up and down by a factor of two.  The
dominant systematic uncertainty comes from the variation of $Q$ in 
$\alpha_\mathrm{S}(Q^2)$ in the matrix element calculation and in the \textsc{Pythia} parton
shower. This variation has a significant effect on the distribution of the
number of jets in \ttV events. 

Systematic uncertainties due to the choice of PDF are evaluated using the
uncertainty sets of the CT10 NLO, MSTW2008 68\% confidence level (CL) NLO and
NNPDF 2.3 NLO~\cite{Ball:2012cx} PDFs following the PDF4LHC
recommendations~\cite{Botje:2011sn}.

The uncertainties on the \ttV modelling are among the dominant ones in the
\FLC\ but they have a negligible impact in all other channels.  Signal
modelling and PDF uncertainties are treated as correlated among channels.

\subsection{Uncertainties on background modelling}
\label{sec:bkg_modeling}

Uncertainties on the background modelling differ significantly among the
channels due to large differences in the background composition.

\paragraph{$Z$ boson background:} \label{sec:syst_zjetsnorm} This dominates in
the \rtwo\ regions of the \OSLC. Four sources of uncertainty are considered:
those associated with the cross section, the $Z$ boson \pt correction, the
scale choice for parton emission, and the choice of generator, evaluated by
comparing the nominal \textsc{Alpgen} sample to a \textsc{Sherpa} sample
generated using \textsc{Sherpa} 1.4.1 with up to three additional partons in
the LO matrix element and the \textsc{CT10} PDF set. 

\paragraph{$\ttbar$ background:} \label{sec:syst_ttbarmodel} This dominates in
the \rone\ regions of the \OSLC. A number of systematic uncertainties affecting
the modelling of the \ttbar process are considered in this channel: those due
to the uncertainty on the cross section which amount to $+5\%$/$-6\%$,  due to
the choice of parton shower and hadronisation model (evaluated by comparing
events produced by \textsc{Powheg} interfaced with \textsc{Pythia} or 
\textsc{Herwig}~\cite{herwig}), due to the choice of generator (evaluated by comparing
a sample generated using \textsc{Madgraph} interfaced with \textsc{Pythia} to the
default \ttbar sample),  and due to the reweighting procedure applied to
correct the {\ttbar} MC modelling.  An additional 50\% normalisation
uncertainty is assigned to \ttbar+heavy-flavour (HF) jets production to account
for limited knowledge of this process.

\paragraph{Single-top background:} \label{sec:ew_norm} This is small and
affects only the \OSLC.  An uncertainty of 6.8\% is assigned to the cross
section for single-top production~\cite{Kidonakis:2010ux}, corresponding to the
theoretical uncertainty on $Wt$ production, the only process contributing to
this final state.  An additional contribution arises from the comparison of
predictions using different schemes to account for interference between $Wt$
and \ttbar.

\paragraph{Diboson background:} In the \TL\ and \SSL\ channels the diboson
background is dominated by $WZ$ production, while in the \FLC\ $ZZ$ production
is dominant.  In the \OSLC\ the diboson background includes $WW$, $WZ$ and $ZZ$
production and the uncertainties are assigned to the sum of these processes.  

In the \TL\ and \SSL\ channels, the normalisation of the $WZ$ background is
treated as a floating parameter in the fit used to extract the \ttV signal. The
uncertainty on the extrapolation of the $WZ$ background estimate from the
control region to signal regions with specific jet and $b$-tag multiplicities
is evaluated by comparing the nominal \textsc{Sherpa} sample to the prediction
of \textsc{Powheg}, as well as by using variations of the simulation
parameters.  The uncertainty amounts to 20\%--35\%.

In the \FL, \TL\ and \SSL\ channels the normalisation of the $ZZ$ background is
treated as a floating parameter in the fit used to extract the \ttV signal.
In the \FLC, several uncertainties on the $ZZ$ background estimate are
considered.  They arise from the extrapolation from the \FLCR\ control region
(corresponding to on-shell $ZZ$ production) to the signal region (with
off-shell $ZZ$ background) and from the extrapolation from the control region
without jets to the signal region with at least one jet. Using data-driven
techniques, these uncertainties are found to be 30\% and 20\%, respectively.
An additional uncertainty of 10\%--30\% is assigned to the normalisation of the
heavy-flavour content of the $ZZ$ background based on a data-to-simulation
comparison of events with one $Z$ boson and additional jets, and cross-checked
with a comparison between different $ZZ$ simulations. 

In the \OSL\ channel, in which the diboson background is small, a 20\%
uncertainty is assigned to the $WZ$ and $ZZ$ background normalisation. This is
estimated from the level of agreement between data and prediction in the \TLCR\
control region. 

\paragraph{\ttH background:} An uncertainty of 12\% is assigned to the \ttH
production cross section~\cite{Dittmaier:2011ti} in all channels.  Additional
uncertainties that affect \ttH kinematics are assigned in the \OSLC\ and are
negligible for the other channels. These uncertainties come from the choice of
factorisation and renormalisation scales, and the functional form of the scale
in \ttH samples. 

\paragraph{$tZ$ and $WtZ$ background:} In the \OSL\ and \TL\ channels, $tZ$ and
$WtZ$ backgrounds are summed and an uncertainty of 20\% is assigned to their
cross section.  An additional uncertainty on the shape of this background is
considered in the \TLC\ for which this background is important. The shape
uncertainty is evaluated by varying the factorisation and renormalisation
scales and $\alpha_\mathrm{S}$ in simulation up and down by a factor of two with
respect to the nominal value, in a correlated manner.  The $\alpha_\mathrm{S}$
variation has the largest effect, ranging from 10\% to 20\% depending on the
number of jets and $b$-jets. 

The $WtZ$ background is important in the \FLC. An uncertainty of 10\% is
assigned to the cross section, coming from the variation of renormalisation and
factorisation scales. An additional uncertainty arises from the modelling of
the additional jet in $WtZ$ events. It is evaluated by varying parameters in
simulation as described above. 

\paragraph{Other prompt lepton backgrounds:} Uncertainties of 20\% are assigned
to the normalisations of the $WH$ and $ZH$ processes, based on calculations
from Ref.~\cite{Heinemeyer:2013tqa}.  An uncertainty of 100\% is considered for
triboson and same-sign $WW$ processes.

\paragraph{Misidentified lepton charge background:} This affects mainly the
\SSLSRA\ and \SSLSRB\ regions.  Uncertainties on it arise from a variety of
statistical and systematic effects.  The main uncertainty comes from the
limited statistical precision in the measurement of charge misidentification
rates.  These are treated as correlated among all \pT and $|\eta|$ bins, but
uncorrelated between the \SSLSRA\ and \SSLSRB\ regions.  This approach is
comparable to treating each binned rate measurement as uncorrelated. Additional
systematic uncertainties arise from the background subtraction in the $Z\to ee$
sample used to measure charge misidentification rates and from the difference
in charge misidentification rates between $Z\to ee$ and \ttbar events. The
latter are evaluated by measuring charge misidentification rates in $Z\to ee$
MC events, assigning them to \ttbar simulated events and comparing the
prediction to the number of true charge-misidentified events in the \ttbar
sample. This uncertainty is found to be 10\%.

\paragraph{Fake lepton background:} This is important in the \SSL, \TL, and
\FL\ channels.  The uncertainty on this background is estimated by propagating
the statistical uncertainty on the measurement of the fake lepton efficiencies.
Additionally, the normalisation of backgrounds in the control regions (from
prompt leptons or charge-misidentified electrons) is varied to estimate a
systematic uncertainty on these efficiencies. The variation assigned (20--25\%)
depends on the composition of each control region, and is chosen to
conservatively cover the largest uncertainty on the backgrounds.  In the \TLC\
an additional uncertainty is considered by measuring the rates of real and fake
leptons in two orthogonal regions, one with three or more jets and the other
with one or two jets.  In the \SSLC, statistical uncertainties are dominant and
no further systematic uncertainties are considered.  All uncertainties
associated with fake leptons are considered to be uncorrelated among analysis
channels and regions.

In the \OSLC\ the fake lepton background is small compared to the other
background contributions. An uncertainty of 50\% is assigned to the fake lepton
yield across all regions in this channel to cover the maximum difference
between yields obtained from the simulation and from \SSL\ events in data.  An
additional uncertainty is assigned to cover the difference in shape of the
distribution of the scalar sum of the transverse momenta of all reconstructed
jets and charged leptons in the simulated and same-sign data events. 

\section{Results}
\label{sec:results}

The observed yields in the 15 signal and 5 control regions are shown
together with the expected numbers of events in \Tab{expected_yields}. 

The production cross sections $\sigma_{\ttW}$ and $\sigma_{\ttZ}$ are
determined simultaneously using a binned maximum-likelihood fit over all
regions and discriminant bins considered in the analysis. The fit is based on
the profile likelihood technique, in which the systematic uncertainties are
treated as nuisance parameters with prior uncertainties that can be constrained
by the fit.  The calculation of confidence intervals and hypothesis testing is
performed using a modified frequentist method as implemented in
RooStats~\cite{RooFit,RooFitManual}. Significance is calculated using the
asymptotic formula of Ref.~\cite{cls_3}.

\begin{figure}[htbp]
\hspace{-3.5ex}
\includegraphics[width=1.06\textwidth]{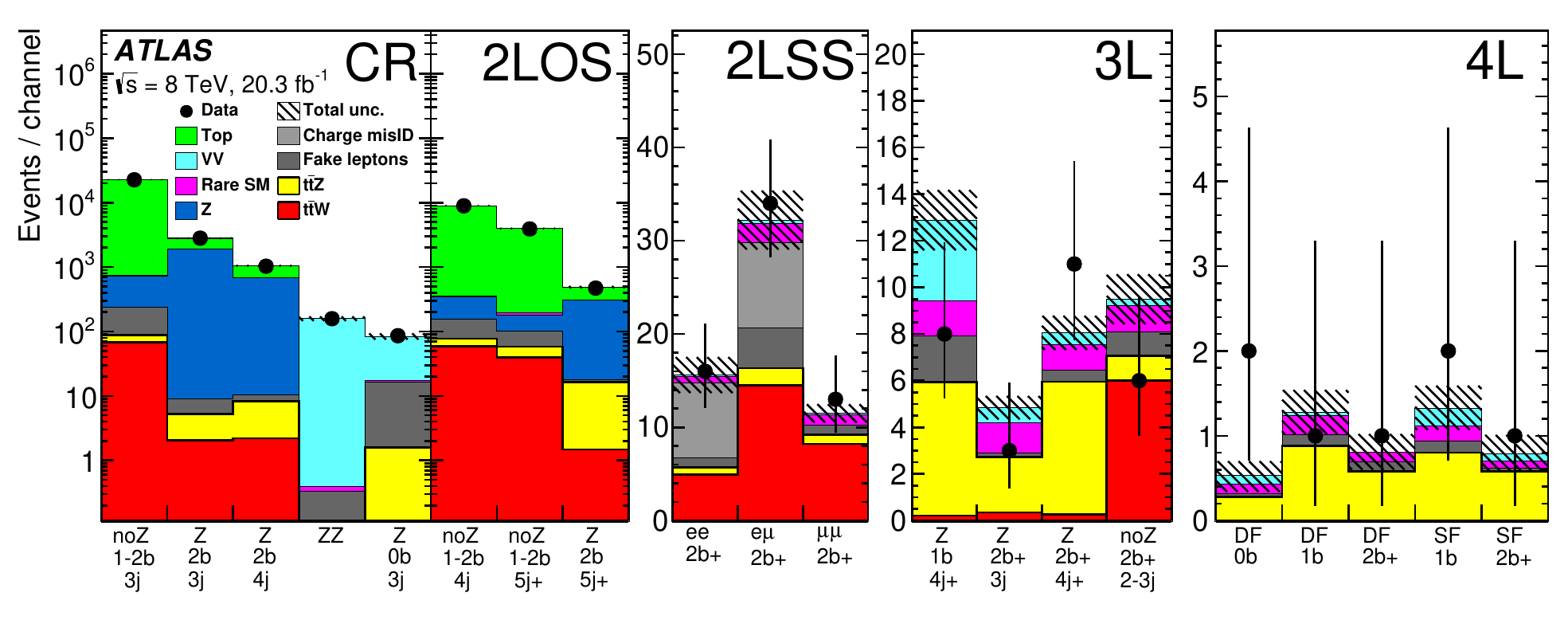}
\caption{Expected yields after the fit compared to data in the five control
regions (CR), used to constrain the \ttbar, $Z$, $ZZ$ and $WZ$ backgrounds, the
three signal regions in the \OSLC\ (2LOS), the three signal regions in the
\SSLC\ (2LSS), the four signal regions in the \TLC\ (3L) and the five signal
regions in the \FLC\ (4L).  In the two dilepton channels the fit also includes
shape information.  The ``Rare SM'' background summarises all other backgrounds
described in \Sect{samples} and mainly consists of the \ttH, $tZ$ and $WtZ$
processes, which are the largest contributions to this background category in
the dilepton, \TL\ and \FL\ channels, respectively. The hatched area
corresponds to the total uncertainty on the predicted yields.}
\label{fig:allyields}
\end{figure}

A summary of the fit to all four channels with their corresponding fit regions
used to measure the \ttW and \ttZ production cross sections is shown in
\Fig{allyields}.  The normalisations of the $WZ$ and $ZZ$ processes are
determined as well, as described in Sections~\ref{sec:wz} and~\ref{sec:zz}.

\Tab{syst} provides a breakdown of the total uncertainties on the measured \ttW
and \ttZ cross sections, determined by fitting each signal individually with
the other fixed to its expected SM value.  For both processes, the precision of
the measurement is dominated by statistical uncertainties.  For the \ttW fit,
the dominant systematic uncertainty source is the modelling of fake leptons and
background processes with misidentified charge.  For the \ttZ fit, the dominant
systematic uncertainty source is the modelling of backgrounds from simulation.

\begin{table}[htbp]
\centering \renewcommand{\arraystretch}{1.2}
\begin{tabular}{|lcc|}
\hline
Uncertainty                 &   $\sigma_{\ttW}$ & $\sigma_{\ttZ}$ \\
\hline
\hline
Luminosity                    &   3.2\% &   4.6\% \\
Reconstructed objects         &   3.7\% &   7.4\% \\
Backgrounds from simulation  &   5.8\% &   8.0\% \\
Fake leptons and charge misID    &   7.5\% &   3.0\% \\
Signal modelling             &   1.8\% &   4.5\%\\
\hline
Total systematic           &   12\% &   13\% \\
Statistical                 &   ${+24\%}\:/\:{-21\%}$ &  ${+30\%}\:/\:{-27\%}$   \\
\hline
\hline
Total                       &   ${+27\%}\:/\:{-24\%}$ &  ${+33\%}\:/\:{-29\%}$  \\
\hline
\end{tabular}
\caption{ Breakdown of uncertainties on the measured cross sections of the \ttW
and \ttZ processes from individual fits. Systematic uncertainties are
symmetrised.}
\label{tab:syst}                                                               
\end{table} 

The sensitivity to the \ttW process is dominated by the \SSLC, while the \ttZ
process is mainly measured in the \TL\ and \FL\ channels.  The result of the
simultaneous fit of the \ttW and \ttZ processes using all four channels is
summarised in \Tab{significances}.  The observed (expected) significance of the
measurements are $5.0\sigma\:(3.2\sigma)$ for the \ttW process and $4.2\sigma\:
(4.5\sigma)$ for the \ttZ process. The background-only hypothesis with neither
\ttZ nor \ttW production is excluded at $7.1\sigma\:(5.9\sigma)$. 

\begin{figure}[htbp]
\centering
\includegraphics[width=0.7\textwidth]{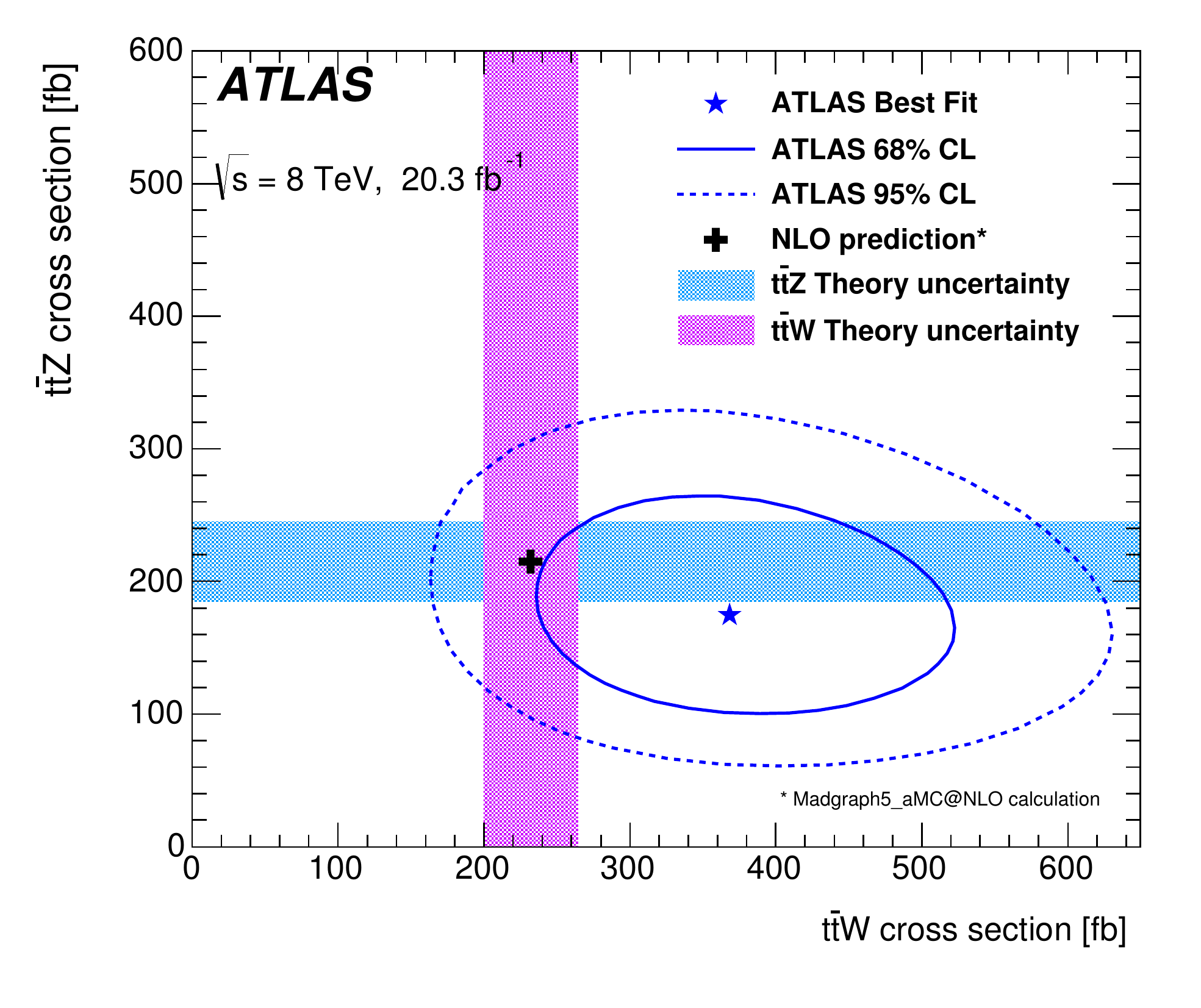}
\caption{The result of the simultaneous fit to
the \ttW and \ttZ cross sections along with the 68\% and 95\% CL
uncertainty contours. The shaded areas correspond to 14\% uncertainty,
which includes renormalisation and factorisation scale uncertainties 
as well as PDF uncertainties including $\alpha_{S}$ variations.}
\label{fig:2D_fit}
\end{figure}

The result of the combined simultaneous fit to the two parameters of interest
is 
\begin{equation}
\sigma_{\ttW} = \vxsttw^{\vxsttwstp}_{\vxsttwstm}\,\mathrm{(stat.)} \pm
\vxsttwsy\,\mathrm{(syst.)\,fb} = \vxsttw^{\vxsttwp}_{\vxsttwm}\,\mathrm{fb}
\end{equation}
and
\begin{equation}
\sigma_{\ttZ} = \vxsttz^{\vxsttzstp}_{\vxsttzstm}\,\mathrm{(stat.)} \pm
\vxsttzsy\,\mathrm{(syst.)\,fb} = \vxsttz^{\vxsttzp}_{\vxsttzm}\,\mathrm{fb}.
\end{equation}
\begin{table}
\centering \renewcommand{\arraystretch}{1.2}
\begin{tabular}{|c|cc|cc|}
\hline 
 & \multicolumn{2}{c|}{\ttW significance} & \multicolumn{2}{c|}{\ttZ
significance} \\
Channel & Expected& Observed & Expected & Observed \\
\hline \hline
2$\ell$OS & 0.4& 0.1& 1.4& 1.1 \\ 
2$\ell$SS & 2.8& 5.0&  - &  -  \\
3$\ell$   & 1.4& 1.0& 3.7& 3.3 \\ 
4$\ell$   &  - &  - & 2.0& 2.4 \\ 
\hline \hline
Combined  & 3.2& 5.0& 4.5& 4.2 \\
\hline 
\end{tabular}
\caption{Expected and observed signal significances for the \ttW and \ttZ
processes determined from the fit to the separate channels and from the
combined fit to all channels. The significance for each signal process is
calculated assuming the null hypothesis for the process in question and
treating the other as a free parameter in the fit. }
\label{tab:significances}
\end{table}
\Fig{2D_fit} provides a comparison of these measurements with NLO QCD
theoretical calculations using \textsc{Madgraph5\_aMC@NLO}.

\section{Conclusion}
\label{sec:conclusion}
Measurements of the production cross sections of a top quark pair in
association with a $W$ or $Z$ boson using \lumi of data collected by the ATLAS
detector in $\sqrt{s} = \SI{8}{\tev}$ $pp$ collisions at the LHC have been
presented. Final states with two, three or four charged leptons are analysed.
From a simultaneous fit to 15 signal regions and 5 control regions, the \ttW
and \ttZ production cross sections are measured to be $\sigma_{\ttW} =
\vxsttw^{\vxsttwp}_{\vxsttwm}\,\mathrm{fb}$ and $\sigma_{\ttZ} =
\vxsttz^{\vxsttzp}_{\vxsttzm}\,\mathrm{fb}$.  The fit to the data considering
both signal processes simultaneously yields significances of $5.0\sigma$ and
$4.2\sigma$ over the background-only hypothesis for the \ttW and \ttZ
processes, respectively.  All measurements are consistent with the NLO QCD
theoretical calculations for \ttW and \ttZ processes. 

\section*{Acknowledgements}

We thank CERN for the very successful operation of the LHC, as well as the
support staff from our institutions without whom ATLAS could not be
operated efficiently.

We acknowledge the support of ANPCyT, Argentina; YerPhI, Armenia; ARC, Australia; BMWFW and FWF, Austria; ANAS, Azerbaijan; SSTC, Belarus; CNPq and FAPESP, Brazil; NSERC, NRC and CFI, Canada; CERN; CONICYT, Chile; CAS, MOST and NSFC, China; COLCIENCIAS, Colombia; MSMT CR, MPO CR and VSC CR, Czech Republic; DNRF, DNSRC and Lundbeck Foundation, Denmark; IN2P3-CNRS, CEA-DSM/IRFU, France; GNSF, Georgia; BMBF, HGF, and MPG, Germany; GSRT, Greece; RGC, Hong Kong SAR, China; ISF, I-CORE and Benoziyo Center, Israel; INFN, Italy; MEXT and JSPS, Japan; CNRST, Morocco; FOM and NWO, Netherlands; RCN, Norway; MNiSW and NCN, Poland; FCT, Portugal; MNE/IFA, Romania; MES of Russia and NRC KI, Russian Federation; JINR; MESTD, Serbia; MSSR, Slovakia; ARRS and MIZ\v{S}, Slovenia; DST/NRF, South Africa; MINECO, Spain; SRC and Wallenberg Foundation, Sweden; SERI, SNSF and Cantons of Bern and Geneva, Switzerland; MOST, Taiwan; TAEK, Turkey; STFC, United Kingdom; DOE and NSF, United States of America. In addition, individual groups and members have received support from BCKDF, the Canada Council, CANARIE, CRC, Compute Canada, FQRNT, and the Ontario Innovation Trust, Canada; EPLANET, ERC, FP7, Horizon 2020 and Marie Skłodowska-Curie Actions, European Union; Investissements d'Avenir Labex and Idex, ANR, Region Auvergne and Fondation Partager le Savoir, France; DFG and AvH Foundation, Germany; Herakleitos, Thales and Aristeia programmes co-financed by EU-ESF and the Greek NSRF; BSF, GIF and Minerva, Israel; BRF, Norway; the Royal Society and Leverhulme Trust, United Kingdom.

The crucial computing support from all WLCG partners is acknowledged
gratefully, in particular from CERN and the ATLAS Tier-1 facilities at
TRIUMF (Canada), NDGF (Denmark, Norway, Sweden), CC-IN2P3 (France),
KIT/GridKA (Germany), INFN-CNAF (Italy), NL-T1 (Netherlands), PIC (Spain),
ASGC (Taiwan), RAL (UK) and BNL (USA) and in the Tier-2 facilities
worldwide.

\clearpage
\printbibliography
\newpage 
\begin{flushleft}
{\Large The ATLAS Collaboration}

\bigskip

G.~Aad$^{\rm 85}$,
B.~Abbott$^{\rm 113}$,
J.~Abdallah$^{\rm 151}$,
O.~Abdinov$^{\rm 11}$,
R.~Aben$^{\rm 107}$,
M.~Abolins$^{\rm 90}$,
O.S.~AbouZeid$^{\rm 158}$,
H.~Abramowicz$^{\rm 153}$,
H.~Abreu$^{\rm 152}$,
R.~Abreu$^{\rm 116}$
% ...
\bigskip
\\
% ...
$^{11}$ Institute of Physics, Azerbaijan Academy of Sciences, Baku, Azerbaijan\\
% ...
$^{85}$ CPPM, Aix-Marseille Universit{\'e} and CNRS/IN2P3, Marseille, France\\
% ...
$^{90}$ Department of Physics and Astronomy, Michigan State University, East Lansing MI, United States of America\\
% ...
$^{107}$ Nikhef National Institute for Subatomic Physics and University of Amsterdam, Amsterdam, Netherlands\\
% ...
$^{113}$ Homer L. Dodge Department of Physics and Astronomy, University of Oklahoma, Norman OK, United States of America\\
% ...
$^{116}$ Center for High Energy Physics, University of Oregon, Eugene OR, United States of America\\
% ...
$^{151}$ Institute of Physics, Academia Sinica, Taipei, Taiwan\\
$^{152}$ Department of Physics, Technion: Israel Institute of Technology, Haifa, Israel\\
$^{153}$ Raymond and Beverly Sackler School of Physics and Astronomy, Tel Aviv University, Tel Aviv, Israel\\
% ...
$^{158}$ Department of Physics, University of Toronto, Toronto ON, Canada\\
% ...
\end{flushleft}

\end{document}